\documentclass[%
reprint,
 amsmath,amssymb,
 aps,
 pre,
]{revtex4-2}

\usepackage{graphicx}
\usepackage{dcolumn}
\usepackage{bm}
\usepackage{multirow}
\newcommand*{\ddt}{\frac{d}{dt}}
\usepackage[usenames,dvipsnames]{color}
\definecolor{dkgreen}{rgb}{0.07, 0.49, 0.12}
\definecolor{mauve}{rgb}{0.58,0,0.82}

\begin{document}

\preprint{APS/123-QED}

\title{A Majority-Vote Model On Multiplex Networks with Community Structure}

\author{Kaiyan Peng}
\affiliation{Department of Mathematics, University of California Los Angeles, Los Angeles, California 90095, USA}
\author{Mason A. Porter}%
\affiliation{%
Department of Mathematics, University of California Los Angeles, Los Angeles, California 90095, USA\\
Santa Fe Institute, 1399 Hyde Park Road, Santa Fe, NM 87501, USA 
}

\date{\today}

\begin{abstract}
We investigate a majority-vote model on two-layer multiplex networks with community structure. In our majority-vote model, the edges on each layer encode one type of social relationship and an individual changes their opinion based on the majority opinions of their neighbors in each layer. To capture the fact that different relationships often have different levels of importance, we introduce a layer-preference parameter, which determines the probability of a node to adopt an opinion when the node's neighborhoods on the two layers have different majority opinions. We construct our networks so that each node is a member of one community on each layer, and we consider situations in which nodes tend to have more connections with nodes from the same community than with nodes from different communities. We study the influence of the layer-preference parameter, the intralayer communities, and interlayer membership correlation on the steady-state behavior of our model using both direct numerical simulations and a mean-field approximation. We find three different types of steady-state behavior: a fully-mixed state, consensus states, and polarized states. We demonstrate that a stronger interlayer community correlation makes polarized steady states reachable for wider ranges of the other model parameters. We also show that different values of the layer-preference parameter result in qualitatively different phase diagrams for the mean opinions at steady states. 

\end{abstract}

\maketitle


\section{Introduction}

The opinions of people who we know often influence our opinions and decisions, and people may change their opinions or behavior to conform to group behavior \cite{asch1951effects}. Researchers have proposed many mathematical models to study such social influence \cite{flache2017}. Many of these models take the form of dynamical processes on networks, where the networks encode contact patterns between individuals \cite{porter2016,newman2018networks}. The simplest type of network is a graph (i.e., a monolayer network), which has a set of nodes to represent individuals and a set of edges to encode pairwise interactions between individuals. In a model of opinion dynamics, the nodes are endowed with one or more opinions, which can change when nodes interact with each other. The modeling of opinion dynamics is subtle and requires many choices; much is unknown about how individuals change their opinions due to the influence of others. Choosing a specific dynamical process to model opinion dynamics entails making major assumptions about the ``microscopic'' behavior of individuals \cite{RevModPhys.81.591, noorazar2020classical, peralta2022opinion}. These models inevitably oversimplify reality, but it is both interesting and relevant to study their population-level and community-level features and to hope that they capture important factors that contribute to trends that we observe in empirical data. 

One opinion model is the \emph{majority-vote model}. In the classical monolayer majority-vote model \cite{de1992isotropic}, each node has one of two opinions. At each time step, a node, which one chooses uniformly at random, adopts the majority opinion of its neighbors with probability $1-f$ and adopts the minority opinion of its neighbors with probability $f$, where $f$ is called the \emph{noise parameter}. Researchers have adapted the majority-vote model to study phenomena such as financial markets \cite{vilela2019majority}, tax evasion \cite{lima2010analysing}, and the effects of filter bubbles (where users on a social-media platform preferentially see content with opinions that are similar to theirs) \cite{vilela2021majority}. 

The classical majority-vote model on a square lattice has a continuous phase transition at a critical value $f_c$ of the noise parameter $f$ \cite{de1992isotropic}. When $f<f_c$, the mean opinion of the population on the lattice is close to $0$ at steady states and (in expectation) the two opinions are distributed uniformly in the population. This regime is called a ``disordered" regime in the physics literature. When $f>f_c$, the system is in an ``ordered" regime in which the absolute value of the {expected} mean opinion is positive and increases with $f$. Researchers have studied the classical majority-vote model on various networks, including Erd\H{o}s--R\'{e}nyi (ER) networks \cite{pereira2005majority, lima2008majority}, Watts--Strogatz (WS) networks \cite{campos2003small, luz2007majority, stone2015majority}, Barab\'{a}si--Albert (BA) networks \cite{lima2006majority}, networks with community structure \cite{huang2015phase}, and several types of lattices \cite{yang2008existence, wu2010majority, santos2011majority,acuna2014critical}. 

In our extension of the majority-vote model, we consider networks with community structure. Community structure is a common feature of social networks \cite{porter2009communities, fortunato2016community}, as individuals tend to have more connections with individuals from the same community than with individuals from different communities. A \emph{stochastic block model} (SBM) \cite{holland1983stochastic, peixoto2019bayesian} is a common generative model for incorporating modules (such as communities) into networks. The simplest type of SBM assigns nodes to disjoint groups (e.g., communities) and samples edges based on group labels. To do this, one defines a matrix $\mathbf P$, where $\mathbf P_{i,j}$ represents the probability that a node in group $i$ is adjacent to a node in group $j$. The \emph{planted-partition model} is a special type of SBM with $\mathbf P_{i,i} = \mathbf P_{\text{in}}$ and $\mathbf P_{i,j} = \mathbf P_\text{out}$ for $i\neq j$.

A large body of research has examined the influence of community structure on a variety of dynamical processes, including the majority-vote model \cite{huang2015phase}, the majority-rule model \cite{lambiotte2007coexistence, lambiotte2007majority}, the Ising model \cite{dasgupta2009phase, bolfe2018phase}, the Sznajd model \cite{ru2008opinion}, the spread of infectious diseases \cite{liu2005epidemic, wu2008community, stegehuis2016epidemic}, a susceptible--infected--recovered (SIR) model \cite{huang2006information}, and the Watts threshold model \cite{nematzadeh2014optimal}. Huang et al.~\cite{huang2015phase} showed that, at a critical value $f_{c_1}$, the majority-vote model on networks with planted community structure undergoes the same continuous disorder--order transition as the majority-vote model does on a square lattice.
 When $f$ is smaller than and close to $f_{c_1}$, all nodes of a network favor one opinion and the distribution of opinions is independent of the community memberships of the nodes. When the noise parameter decreases further, the system in Ref.~\cite{huang2015phase} undergoes a discontinuous phase transition at a second critical value $f_{c_2}$ under certain conditions. When $f< f_{c_2}$, the mean opinions of the different communities can have different signs, which means that the different communities favor different opinions. Researchers have also observed the presence of clusters of nodes with different opinions when studying the Ising model \cite{suchecki2009bistable, dasgupta2009phase, bolfe2018phase} and the majority-rule model \cite{lambiotte2007coexistence, lambiotte2007majority} on networks with community structure.

Social relationships in real life are multifaceted \cite{kivela2014multilayer}, and it is important to account for this complexity in the study of opinion dynamics. One approach is to consider opinion dynamics on \emph{multilayer networks} \cite{kivela2014multilayer, aleta2019multilayer, de2022multilayer}. A multilayer network has layers in addition to nodes and edges.  A node can exist in one or multiple layers, and one can use different layers to encode different types of interactions.The instantiation of an entity in a layer is called a \emph{state node}. The set of all state nodes of the same entity is a \emph{physical node}. In the most general form, any two state nodes can be adjacent to each other. The \emph{interlayer edges} (i.e., edges that connect nodes from different layers) introduce the possibility of structural correlations between the layers. This cannot occur in monolayer networks. In our paper, we focus on a special type of multilayer network, called a \emph{multiplex network}, in which interlayer edges can only connect state nodes that correspond to the same physical node. Edges in different layers (i.e., \emph{intralayer edges}) can have different connectivity patterns.

Each layer in a multiplex network can encode a different type of relationship. In this scenario, it is reasonable to assume that all nodes that correspond to a physical node share the same opinion. Different relationships differ in how they are formed and maintained, and they have different levels of importance to an individual. Consequently, the opinions from disparate sources may spread following different rules. Another possibility is that the layers of a multiplex network represent different topics. In this case, the state nodes that correspond to the same physical node can have distinct opinions. Many dynamical processes on multilayer networks cannot be reduced to equivalent processes on monolayer networks that one obtains by aggregating multiplex networks \cite{diakonova2016irreducibility}. Researchers have studied several types of opinion dynamics on multilayer networks. Examples include the majority-vote model \cite{ krawiecki2018majority, choi2019majority}, the voter model \cite{diakonova2016irreducibility}, the $q$-voter model \cite{chmiel2017tricriticality, gradowski2020pair}, the Watts threshold model~\cite{lee2014threshold}, and the Deffuant--Weisbuch (DW) model \cite{nguyen2018opinion}. 

In the present paper, we study steady-state behavior in a majority-vote model on multiplex networks with community structure. For simplicity, we consider two-layer networks. To incorporate community structure into our multiplex networks, we construct each layer using an SBM. We assume that each layer represents one social relationship and that each physical node holds one opinion. When the two layers have the same majority opinion, the update rules are the same as in the monolayer majority-vote model. Previous extensions of the majority-vote model to multiplex networks \cite{krawiecki2018majority, choi2019majority} (which also considered two-layer networks) treated all layers equally, so that when the majority opinions differ in different layers, a focal node either uniformly randomly adopts an opinion or maintains its current opinion. However, in real life, individuals can attach different levels of importance to different relationships or favor some social relationships over others. We model this situation by introducing a layer-preference parameter that controls the probability of adopting each opinion when the two layers in a network do not have a common majority opinion. We study the influence of the layer-preference parameter, the intralayer communities, and interlayer membership correlations on the steady-state behavior of our model with both direct numerical simulations and a mean-field approximation.

Our paper proceeds as follows. In Section \ref{sec: mv-model}, we introduce our majority-vote model and discuss the structure of the multiplex networks on which we consider its dynamics. We derive a mean-field approximation of our model in Section \ref{sec: mv-mf}; this yields a coupled system of ODEs. In Section \ref{sec: mv-steady-state}, we conduct linear stability analysis of three steady-state behaviors that we observe in our mean-field system. In Section \ref{sec: mv-exp}, we compare the results of direct numerical simulations with those of our mean-field approximation and explore the influence of the model parameters on the locations of our model's phase transitions. We conclude in Section \ref{sec: mv-conclusion}.


\section{Our Model}\label{sec: mv-model}

We consider a two-layer multiplex network of $N$ physical nodes, and we suppose that all physical nodes are present in both layers. Each physical node is associated with one of two opinions, which we label as $1$ and $-1$. The two layers encode different types of relationships, which can be online and offline relationships, LinkedIn and Facebook friendship relationships, or something else. We label the two layers as layer 1 and layer 2. We introduce our majority-vote model in Section \ref{sec: mv-dynamics} and discuss the network structure in Section \ref{sec: mv-network}.


\subsection{Majority-vote dynamics on multiplex networks}\label{sec: mv-dynamics}

Our extension of the majority-vote model emphasizes layer heterogeneity. Suppose that our system updates at discrete time steps. At each time step, we choose a physical node uniformly at random. We denote its opinion by $O_A$ and denote the other opinion by $O_B$. This focal node conducts a local survey of its neighbors' opinions on each layer. There are four possible situations, which we summarize in Table \ref{table: mv}. If the opinion $O_A$ is the common majority opinion in both layers, the chosen node flips its opinion with probability $f_1$. If $O_A$ is the majority opinion in layer 1 but not layer 2, the node changes its opinion and adopts opinion $O_B$ with probability $f_2$. If $O_A$ is the majority opinion in layer 2 but not in layer 1, it changes its opinion with probability $1-f_2$. Finally, if $O_A$ is not the majority opinion in either layer, the focal node changes its opinion with probability $1 - f_1$. The parameter $f_1$ is a natural extension of the noise parameter in the classical majority-vote model, so we refer to it as the ``noise parameter'' in this paper. We only consider $f_1 < 0.5$, as nodes are more likely to follow the majority opinion. We introduce the parameter $f_2$ to model the situation that neighbors in one layer are more influential than neighbors in the other layer. We refer to this parameter as the ``layer-preference parameter''. Without loss of generality, we assume that $f_2 \leq 0.5$, which implies that neighbors in layer 1 are at least as persuasive as neighbors in layer 2. When $f_2 = 0.5$, our model is similar to the model that was studied in Ref.~\cite{krawiecki2018majority}; when $f_2 = 0$, our model is similar to the ``AND'' model that was studied in Ref.~\cite{choi2019majority}. We also assume that $f_2 > f_1$, so a node is more likely to adopt an opinion that is the common majority opinion in both layers than an opinion that is the majority opinion in only one of the layers.

\begin{table}[]
\centering
\caption{{The probability that a focal node changes its opinion. We denote the original opinion of the focal node by $O_A$ and denote the other opinion by $O_B$. The first column indicates the majority opinion in each layer. We use $\bar O_A$ to indicate that $O_A$ is not the majority opinion in a layer. The second column gives the probability that the focal node adopts opinion $O_B$.}
}\label{table: mv}
\begin{ruledtabular}
\begin{tabular}{cc|c}
\multicolumn{2}{c|}{Majority opinions} & \multirow{2}{*}{Opinion-flip probability~~~~~~ } \\
Layer 1            & Layer 2           &                                           \\ \hline
$O_A$                & $O_A$               & $f_1$                                     \\
$O_A$                & $\bar O_A$              & $f_2$                                     \\
$\bar O_A$               & $O_A$               & $1-f_2$                                   \\
$\bar O_A$               & $\bar O_A$              & $1-f_1$                                   \\ 
\end{tabular}
\end{ruledtabular}
\end{table}


\subsection{Multiplex networks with community structure}\label{sec: mv-network}

To incorporate community structure into our multiplex networks, we construct each layer using an undirected SBM. For simplicity, we assume that each layer has two communities and that each community consists of $N/2$ nodes (and that $N$ is even). For physical nodes, there are 4 different community labels, which we denote by $\mathbf g = (g_1, g_2)$, where $g_l\in \{1, 2\}$ is the label in layer $l$. To control the correlation of community assignments across layers, we consider the probability $\nu$ that a physical node belongs to community $g$ (with $g \in \{1,2\}$) in both layers. We calculate that $\mathbb E[\vert C_{(1,1)}\vert] = \mathbb E[\vert C_{(2,2)} \vert ]= \nu N/2$ and $\mathbb E[\vert C_{(1,2)}\vert] = \mathbb E[\vert C_{(2,1)}\vert] = (1-\nu )N/2$, where $C_{(g_1,g_2)}$ denotes the set of physical nodes with community label $(g_1, g_2)$ and $\mathbb E[\vert \cdot \vert]$ denotes the expected cardinality of a set. In our simulations, we independently assign each physical node to community $C_{\mathbf g}$ with probability $\mathbb E[\vert C_{\mathbf g} \vert ]$/N. Without loss of generality, we assume that $\nu \geq 0.5$. One can exchange community labels in one of the two layers to obtain an equivalent model with $\nu < 0.5$. After we determine community assignments, we generate intralayer edges independently in each layer using an SBM. To do this, in each layer $l$, we connect two state nodes that belong to the same community with probability $p_{\text{in},l}$ and we connect two state nodes that belong to different communities with probability $p_{\text{out},l}$, where $0 < p_{\text{out},l} < p_{\text{in},l}$. We measure the strength of intralayer communities with the parameter $\mu_l = p_{\text{out},l}/(p_{\text{out},l} + p_{\text{in},l}) \in(0, 0.5)$. A smaller value of $\mu_l$ means that a larger fraction of edges connect nodes that are in the same community. We use a simplistic network structure for convenience. More realistic network structures can include more layers and more communities; each layer can have different numbers of communities of different sizes and different edge densities. To use such a general setting, one can impose interlayer community correlations and construct intralayer communities following Ref.~\cite{bazzi2020framework}. One can also study the influence of more sophisticated community structures, such as ones with overlapping communities.


\section{A mean-field approximation}\label{sec: mv-mf}

In this section, we derive a mean-field approximation of our majority-vote model and study the dynamics of the approximation.
We group nodes with the same degrees. We denote the degree of a physical node by $\mathbf k = (k_1, k_2)$, where $k_l$ is the node's degree in layer $l$. We also group nodes based on their community assignments. Let $q_{\mathbf k,\mathbf g}$ denote the probability that a uniformly randomly chosen node with degree $\mathbf k$ in community $C_{\mathbf g}$ holds opinion $1$. The probability $q_{\mathbf k,\mathbf g}$ satisfies
\begin{eqnarray}
    \ddt q_{\mathbf k,\mathbf g} =  (1- q_{\mathbf k,\mathbf g})P_{\mathbf k,\mathbf g}^{-1 \rightarrow 1} -  q_{\mathbf k,\mathbf g}P_{\mathbf k,\mathbf g}^{1 \rightarrow -1}\,, \label{equ: mv-eq1}
\end{eqnarray}
where $P_{\mathbf k,\mathbf g}^{O \rightarrow -O}$ is the probability that a node changes its opinion to $-O$ from opinion $O \in \{1,-1\}$. The probability $P_{\mathbf k,\mathbf g}^{O\rightarrow -O}$ depends on the opinion distribution of a focal node's neighbors, and this opinion distribution depends implicitly on the opinion of the focal node. However, as a simplification, we assume that the states of the neighbors and the state of the focal node are independent when updating the state of the focal node. This assumption was called ``absence of dynamical correlations'' in Ref.~\cite{gleeson2012accuracy}. Additionally, we only consider the case that both layers have majority opinions; we ignore all other situations. In our simulations, we examine several opinion-update probabilities for when one or both layers do not have a majority opinion, but we do not observe meaningful differences between the steady-state behaviors of these different choices. Therefore, we do not present these results in the present paper and we claim that ignoring the case 
of an equal number of neighbors supporting each opinion in a layer has minimal influence on our mean-field expressions. With these simplifications, we define $P_{\mathbf k,\mathbf g}^{\rightarrow O}$ as the probability that the focal node adopts opinion $O$ regardless of its current state. The probability $P_{\mathbf k,\mathbf g}^{\rightarrow O}$ satisfies $P_{\mathbf k,\mathbf g}^{\rightarrow -1} = 1 - P_{\mathbf k,\mathbf g}^{\rightarrow 1}$. 
Let $ \bar q_{\mathbf g} = \sum_{\mathbf k }\mathbb P(\mathbf k)  q_{\mathbf k,\mathbf g}$, where $\mathbb P(\mathbf k)$ denotes the degree distribution of the physical nodes. Based on equation \eqref{equ: mv-eq1}, we have
\begin{align}
    \ddt \bar q_{\mathbf g} =   -  \bar q_{\mathbf g} + \sum_{\mathbf k}\mathbb P(\mathbf k) P_{\mathbf k,\mathbf g}^{ \rightarrow 1}\,. \label{equ: mv-eq2}
\end{align}

We seek to expand $P_{\mathbf k,\mathbf g}^{ \rightarrow 1}$. Let $\xi^{(l)}_{k,g}$ denote the probability that the majority opinion of the neighbors is 1 for a focal state node in layer $l$ with degree $k$ and community label $g$. We assume that the states of the neighbors in the two layers are independent, so
\begin{align}
     P_{\mathbf k,\mathbf g}^{ \rightarrow 1} & =  (1-f_1)\xi^{(1)}_{k_1,g_1}\xi^{(2)}_{k_2,g_2} + f_1 (1-\xi^{(1)}_{k_1,g_1})(1-\xi^{(2)}_{k_2,g_2}) \notag\\
    & ~~~~+ (1-f_2)\xi^{(1)}_{k_1,g_1}(1-\xi^{(2)}_{k_2,g_2}) + f_2 (1-\xi^{(1)}_{k_1,g_1})\xi^{(2)}_{k_2,g_2} \notag\\
     & =  f_1 + (1-f_2-f_1)\xi^{(1)}_{k_1,g_1} +
    (f_2-f_1)\xi^{(2)}_{k_2,g_2}
    \,.\label{equ: mv-eq3}
\end{align}
The probability $\xi^{(l)}_{k,g}$ is
\begin{align}
    \xi^{(l)}_{k,g} = \sum_{n=\lceil k/2 \rceil}^{k}\left(1-\frac{1}{2}\delta_{n,k/2}\right){{k}\choose n }(Q^{(l)}_{g})^n(1-Q^{(l)}_{g})^{k-n}\,,\label{equ: mv-eq4}
\end{align}
where $\delta$ is the Kronecker delta and $Q^{(l)}_{g}$ denotes the probability that a uniformly randomly chosen neighbor of a state node with community label $g$ in layer $l$ holds opinion 1.
To close the system, we write $Q^{(l)}_{g}$ in terms of $\bar q_{\mathbf g}$ as follows:
\begin{widetext}
\begin{equation}
    \begin{split}
        Q^{(1)}_{1} & = (1-\mu_1)\nu \bar q_{(1,1)} 
        + (1-\mu_1)(1-\nu) \bar q_{(1,2)} \
         + \mu_1(1-\nu)\bar q_{(2,1)} + \mu_1\nu\bar  q_{(2,2)}\,,\\
        Q^{(1)}_{2} & = \mu_1\nu\bar  q_{(1,1)} + \mu_1(1-\nu) \bar q_{(1,2)} 
       + (1-\mu_1)(1-\nu)\bar q_{(2,1)} + (1-\mu_1)\nu\bar  q_{(2,2)}\,,\\
        Q^{(2)}_{1} & = (1-\mu_2)\nu \bar q_{(1,1)} + \mu_2(1-\nu)\bar  q_{(1,2)} 
         + (1-\mu_2)(1-\nu)\bar q_{(2,1)} + \mu_2\nu \bar q_{(2,2)}\,,\\
        Q^{(2)}_{2} & = \mu_2\nu \bar q_{(1,1)} + (1-\mu_2)(1-\nu)\bar  q_{(1,2)} 
         + \mu_2(1-\nu)\bar q_{(2,1)} + (1-\mu_2)\nu \bar q_{(2,2)}\,.\label{equ: mv-eq5}
    \end{split}
\end{equation}
\end{widetext}
Equations \eqref{equ: mv-eq2}--\eqref{equ: mv-eq5} are a closed system that approximates the time evolution of $\bar q_{\mathbf g}$.

The binomial distribution in equation \eqref{equ: mv-eq4} is expensive to compute for large values of $k$. Therefore, we approximate it with a normal distribution using the central limit theorem. In our subsequent analysis and experiments, we use the following approximation:
\begin{equation}
    \xi^{(l)}_{k,g} \approx \frac{1}{2} + \frac{1}{2}\mathrm{erf}\left(\sqrt{2k}\left(Q^{(l)}_{g}-\frac{1}{2}\right)\right)\,,\label{equ: mv-eq6}
\end{equation}
where erf$(z) = \frac{2}{\sqrt{\pi}}\int_0^z e^{-t^2}dt$ is the error function. 


\section{Steady states}\label{sec: mv-steady-state}

By numerically solving the ODE system (\ref{equ: mv-eq2}, \ref{equ: mv-eq3}, \ref{equ: mv-eq5}, \ref{equ: mv-eq6}), we find that there are three types of steady states; these states depend on the choices of parameters and on the initial conditions. {In this section, we derive expressions for the steady-state solutions and conditions for the linear stability of these solutions using our mean-field approximation (\ref{equ: mv-eq2}, \ref{equ: mv-eq3}, \ref{equ: mv-eq5}, \ref{equ: mv-eq6}).} In our linear-stability calculations, we use the Perron--Frobenius theorem and consider the largest positive eigenvalue of the Jacobian matrix.

Let $\bar m_{(g_1, g_2)} = 2\bar q_{(g_1, g_2)} - 1$ denote the {expected} mean opinion of community $C_{(g_1, g_2)}$, and let $\bar m = \sum_{g_1, g_2} \mathbb E[\vert C_{g_1, g_2}\vert ] \bar m_{(g_1, g_2)}/N$ denote the {expected} mean opinion of the entire population. One steady-state solution is $\bar q_{(g_1, g_2)} = \frac{1}{2}$ for all $g_1$ and $g_2$. In this solution, the two opinions are uniformly randomly distributed in the population and the mean opinion $\bar m_{(g_1, g_2)}$ of each community is $0$. We refer to this solution as the ``fully-mixed steady state''. The four communities in the other two types of steady-state solutions have preferences either for the same opinion or for different opinions. In one case, the solution $\bar q_{(g_1, g_2)}$ deviates from $\frac{1}{2}$ and has equal values for all $g_1$ and $g_2$; such states are ``consensus steady states''. In the other case, $\bar q_{(g_1, g_2)} - \frac{1}{2}$ do not all have the same sign; these are ``polarized steady states''. 


\subsection{The fully-mixed steady state}

We start with the fully-mixed steady-state solution. The Jacobian matrix of equation \eqref{equ: mv-eq2} at $\bar q_{(g_1, g_2)} = \frac{1}{2}$ for all $g_1,\,g_2$ is 
\begin{widetext}
\begin{align}
    \mathbb J_{\text{full}} & =  \frac{\sqrt2}{\sqrt\pi}(1-f_2-f_1)\langle\sqrt k\rangle^{(1)} \notag
     \times 
    \begin{bmatrix}
    (1-\mu_1)\nu & (1-\mu_1)(1-\nu) & \mu_1(1-\nu) & \mu_1\nu\\
     (1-\mu_1)\nu  & (1-\mu_1)(1-\nu) & \mu_1(1-\nu) & \mu_1\nu\\
    \mu_1\nu & \mu_1(1-\nu) & (1-1\mu_1)(1-\nu) &(1-\mu_1) \nu\\
   \mu_1\nu & \mu_1(1-\nu) & (1-1\mu_1)(1-\nu) & (1-\mu_1)\nu
    \end{bmatrix}\\
    &\quad +  \frac{\sqrt2}{\sqrt\pi}(f_2-f_1)\langle\sqrt k\rangle^{(2)} \times
    \begin{bmatrix}
    (1-\mu_2)\nu & \mu_2(1-\nu) & (1-\mu_2)(1-\nu) & \mu_2\nu\\
    \mu_2\nu & (1-\mu_2)(1-\nu) & \mu_2(1-\nu) & (1-\mu_2)\nu\\
    (1-\mu_2)\nu & \mu_2(1-\nu) & (1-\mu_2)(1-\nu) & \mu_2\nu\\
     \mu_2\nu & (1-\mu_2)(1-\nu) & \mu_2(1-\nu) & (1-\mu_2)\nu
    \end{bmatrix} - I\,,\label{eq: fm}
\end{align}
\end{widetext}
where $\langle\sqrt{k} \rangle^{(l)}$ denotes the mean of the square root of the degree in layer $l$. Because $0.5\geq f_2>f_1$, the largest eigenvalue of $\mathbb J_{\text{full}}$ is 
\begin{equation}
   \lambda_{\text{full}} =  - 1 + \frac{\sqrt2}{\sqrt \pi} (1-f_2-f_1) \langle \sqrt{k}\rangle^{(1)} + \frac{\sqrt2}{\sqrt\pi}(f_2-f_1)\langle \sqrt{k}\rangle^{(2)}\,. \label{eq: fm-eig}
\end{equation}
Therefore, the fully-mixed steady state $\bar q_{(g_1, g_2)} = \frac{1}{2}$ (with $g_1,\,g_2\in\{1,2\}$) is linearly stable if $\lambda_{\text{full}} < 0$. This condition is independent of $\mu_1$, $\mu_2$, and $\nu$. This result matches what the result that Huang et al.~\cite{huang2015phase} obtained for the majority-vote model on monolayer networks with community structure. Increasing $f_1$ always decreases $\lambda_{\text{full}}$; this helps stabilize the fully-mixed solution. By contrast, the effect of $f_2$ depends on the difference between the edge densities of the two layers. Recall that $f_2$ measures the tendency for nodes to favor the majority opinion in layer $1$ over that in layer $2$ when their majority opinions differ. A smaller value of $f_2$ implies that nodes have a larger preference for layer $1$. When layer $1$ has fewer (respectively, more) edges, which is equivalent to the condition that layer $1$ has a smaller (respectively, larger) value of $\langle k\rangle$, decreasing (respectively, increasing) $f_2$ helps stabilize the fully-mixed state. 


\subsection{Consensus steady states}\label{mv: consensus}

We suppose that each community has the same mean opinion. To do this, we examine steady-state solutions of the system (\ref{equ: mv-eq2}, \ref{equ: mv-eq3}, \ref{equ: mv-eq5}, \ref{equ: mv-eq6}) of the form 
\begin{equation}
         \bar q_{\mathbf g} = \frac{1}{2} + \epsilon\,\quad \text{for all }\mathbf g\label{equ: mv-eq11}
\end{equation}
for some $\epsilon\in[-\frac{1}{2}, 0)\cup(0, \frac{1}{2}]$. Consequently, the mean opinions $\bar m_{\textbf{g}} = \bar m = 2\epsilon \neq 0$. 
We set the right-hand sides of equations \eqref{equ: mv-eq2} to 0 and insert the ansatz \eqref{equ: mv-eq11}. The parameter $\epsilon$ satisfies 
\begin{eqnarray}
    \epsilon & = \frac{1}{2}(1-f_2-f_1)\left (\sum_{k_1}\mathbb P^{(1)}(k_1)\mathrm{erf}(\sqrt{2k_1}\epsilon) \right ) \notag \\
    &\quad + \frac{1}{2}(f_2-f_1)\left (\sum_{k_2}\mathbb P^{(2)}(k_2)\mathrm{erf}(\sqrt{2k_2}\epsilon)\right )\,,\label{equ: mv-eq12}
\end{eqnarray}
where $\mathbb P^{(l)}(k)$ denotes the degree distribution of the state nodes in layer $l$. 
The error function is an S-shaped odd function and is bounded by $-1$ and $1$. Therefore, equation \eqref{equ: mv-eq12} has a solution $\epsilon \in [-\frac{1}{2}, 0) \cup (0, \frac{1}{2}]$ if and only if the value on the right-hand side of \eqref{equ: mv-eq12} is smaller than $\frac{1}{2}$ at $\epsilon=\frac{1}{2}$ and its derivative at $\epsilon=0$ is larger than $1$. The former condition is always satisfied. The latter condition is equivalent to
\begin{equation}
    1 < \frac{\sqrt2}{\sqrt \pi} (1-f_2-f_1)\langle \sqrt{k}\rangle^{(1)} + \frac{\sqrt2}{\sqrt \pi} (f_2-f_1)\langle \sqrt{k}\rangle^{(2)} \,,\label{equ: mv-stability-first}
\end{equation}
which is also the condition for the disordered steady state to be unstable. Suppose that equation \eqref{equ: mv-eq12} has a non-zero solution $\epsilon^* \in [-\frac{1}{2},\frac{1}{2}]$. The solution $\bar q_{\mathbf g} = \frac{1}{2} + \epsilon^*$ is always linearly stable because the largest eigenvalue of the Jacobian matrix of equation \eqref{equ: mv-eq2} at these values of $\bar q_{\mathbf g}$
equals the derivative of the right-hand side of equation \eqref{equ: mv-eq12} minus $1$. The derivative is always smaller than $1$. Therefore, for a fixed $f_2$, when $f_1$ is smaller than a critical value, consensus steady states are linearly stable. Additionally, the mean opinions $\bar m_{\mathbf g}$ at a consensus steady state and the critical value of $f_1$ are independent of the values of $\mu_1$, $\mu_2$, and $\nu$.


\subsection{Polarized steady states}

Suppose that the two communities in one layer have different mean opinions at steady state. Because of the symmetry between the two opinions and the symmetry between the two communities, we expect that the steady-state solutions also have a symmetric structure. Let
\begin{align*}
    \bar q_{(1,1)} &= \frac{1}{2} + \epsilon_1\,, \qquad  \bar q_{(1,2)} = \frac{1}{2} + \epsilon_2\,,\\
    \bar q_{(2,1)} &= \frac{1}{2} - \epsilon_2\,, \qquad  \bar q_{(2,2)} = \frac{1}{2} - \epsilon_1\,.
\end{align*}
Inserting this ansatz into the steady state of the system (\ref{equ: mv-eq2}, \ref{equ: mv-eq3}, \ref{equ: mv-eq5}, \ref{equ: mv-eq6}) yields
\begin{equation}
    \begin{split}
        \epsilon_1 & = \frac{1}{2}(1-f_2-f_1)\left(\sum_{k_1}\mathbb P^{(1)}(k_1) \mathrm{erf}(\sqrt{2k_1}y_{1})\right) \\
        &\quad + \frac{1}{2}(f_2-f_1)\left(\sum_{k_2}\mathbb P^{(2)}(k_2) \mathrm{erf}(\sqrt{2k_2}y_{2})\right)\,,\\
         \epsilon_2 & = \frac{1}{2}(1-f_2-f_1)\left(\sum_{k_1}\mathbb P^{(1)}(k_1) \mathrm{erf}(\sqrt{2k_1}y_{1})\right) \\
         &\quad - \frac{1}{2}(f_2-f_1)\left(\sum_{k_2}\mathbb P^{(2)}(k_2) \mathrm{erf}(\sqrt{2k_2}y_{2})\right)
        \,,
        \end{split}\label{equ: mv-polarization_fixed_point}
\end{equation}
where 
\begin{equation}
    \begin{split}
        y_1 = (1-2\mu_1)\nu\epsilon_1 + (1-2\mu_1)(1-\nu)\epsilon_2\,,\\
        y_2 = (1-2\mu_2)\nu\epsilon_1 - (1-2\mu_2)(1-\nu)\epsilon_2\,.
    \end{split}\label{equ: mv-polarization_y}
\end{equation} 
One can solve equations \eqref{equ: mv-polarization_fixed_point}--\eqref{equ: mv-polarization_y} for $\epsilon_1$ and $\epsilon_2$ using fixed-point iterations and then determine the linear stability of the polarized steady states by evaluating the largest eigenvalue of the Jacobian matrix. This eigenvalue is
{\small
\begin{equation}
\begin{split}
        \lambda_{\text{polarized}}  & =  \frac{\sqrt{2}}{\sqrt{\pi}}(1-f_2-f_1)\left(\sum_{k_1}\mathbb P^{(1)}(k_1)\sqrt{k_1} e^{-(\sqrt{2k_1}y_{1})^2}\right) \\
        &\quad 
        + \frac{\sqrt{2}}{\sqrt{\pi}}(f_2-f_1)\left(\sum_{k_2}\mathbb P^{(2)}(k_2)\sqrt{k_2} e^{-(\sqrt{2k_2}y_{2})^2}\right) - 1
        \,.\label{equ: mv-polarization_condition}
\end{split}
\end{equation}
}

Consider the special case in which we construct the two layers of a multiplex network from the same random-network ensemble. Specifically, we take $\mu_1 = \mu_2 =\mu$ and $\mathbb P^{(1)}(k) = \mathbb P^{(2)}(k)=\mathbb P(k)$. Additionally, we let $f_2 = 0.5$. Because of the symmetry between the two layers, $\epsilon_2 = 0$ and 
\begin{equation}
    \epsilon_1 = \left(\frac{1}{2}-f_1\right)\sum_{k}\mathbb P(k) \mathrm{erf}\left(\sqrt{2k}(1-2\mu)\nu\epsilon_1\right)\,.\label{equ: mv-eq7}
\end{equation}
As we discussed in Section \ref{mv: consensus}, equation \eqref{equ: mv-eq7} has a nontrivial solution if the derivative of its right-hand side at $\epsilon_1=0$ is larger than 1. Equivalently, there is a nontrivial solution if 
\begin{align}
    f_1 < \frac{1}{2} - \frac{\sqrt{\pi}}{2\sqrt{2}(1-2\mu)\nu\langle k^{1/2}\rangle}\,.\label{equ: mv-eq8}
\end{align}
Suppose that condition \eqref{equ: mv-eq8} is satisfied, We denote the solution of \eqref{equ: mv-eq7} by $\epsilon_1^*$. The largest eigenvalue of the Jacobian matrix is $-1  +  \frac{2\sqrt2}{\sqrt\pi}(\frac{1}{2}-f)\sum_{k}\mathbb P(k)\sqrt ke^{-(\sqrt{2k}(1-2\mu)\nu\epsilon_1^*)^2}$. Therefore, the steady-state solution is linearly stable if and only if
\begin{equation}
    \begin{split}
        \epsilon_1 & = \left(\frac{1}{2}-f_1\right)\sum_{k}\mathbb P(k) \mathrm{erf}\left(\sqrt{2k}(1-2\mu)\nu\epsilon_1\right)\,,\\
        1 & >  \frac{2\sqrt2}{\sqrt\pi}\left(\frac{1}{2}-f_1\right)\sum_{k}\mathbb P(k)\sqrt ke^{-\left(\sqrt{2k}(1-2\mu)\nu\epsilon_1\right)^2}\,.
    \end{split}\label{equ: mv-stability-2}
\end{equation}
In this special case, consensus steady states are linearly stable if and only if $f_1<\frac{1}{2} - \frac{\sqrt{\pi}}{2\sqrt{2}\langle k^{1/2}\rangle}$, which is larger than the right-hand side of equation \eqref{equ: mv-eq8} because $\mu\in(0, 0.5)$ and $\nu\in[0.5, 1]$. Importantly, because equation \eqref{equ: mv-eq8} is only a necessary condition for polarized steady states to be linearly stable, the stability region of the polarized steady states is a subset of the stability region of the consensus steady states. This result is consistent with and generalizes previous results for the monolayer majority-vote model \cite{huang2015phase}. 


\section{Numerical experiments}\label{sec: mv-exp}

We now compare the results of computational simulations with our analytical results (see our code at \cite{code}). In all of our simulations, we generate multiplex networks using the network model in Section \ref{sec: mv-network}. We sample the two layers from the same network ensemble. We set the mean degrees to $\langle k \rangle = 60$ in each layer and let $\mu_1 = \mu_2 = \mu = 0.25$. Consequently, the degree distribution of state nodes in each layer approximately follows a Poisson distribution with mean $60$ and $(p_{\text{in}}, p_{\text{out}}) = \left(\frac{2(1-\mu) \langle k \rangle}{N}, \frac{2\mu \langle k \rangle}{N}\right)$. We run each simulation for $200\,N$ time steps. We consider a steady state to be stable (and use the word ``stable'' on its own) if we observe it in our simulations. In our simulations, when we assign the initial opinions uniformly at random, the system can reach either consensus steady states or polarized steady states if the system is in the regime in which both types of steady states are stable. However, because (1) the linear-stability region of the consensus steady states contains the linear-stability region of polarized steady states in our mean-field model and (2) we are more interested in locating the boundary at which polarized steady states lose stability, we want more simulations to reach polarized steady states. To achieve this, we assign the initial opinions based on the community assignments of the nodes. We assign opinion $1$ to all physical nodes in community $1$ in layer $1$ and assign opinion $-1$ to the remaining physical nodes. By maximizing $\vert \bar m_{(1,1)} - \bar m_{(2,2)}\vert$ and $\vert \bar m_{(1,2)} - \bar m_{(2,1)}\vert$, we expect that the initial state of the system is closer in the state space to a polarized steady state than to a consensus steady state. Consequently, we expect that it is more difficult for stochasticity in our numerical simulations to pull the system towards a consensus steady state than towards a polarized steady state. In our simulations, this choice of initial opinions always yields polarized steady states when the parameters are inside the polarized regime and are sufficiently far away from its stability boundary. We initialize the opinions of the nodes based on their community assignments in all subsequent simulations.

For both our simulations and our mean-field approximation, we calculate the overall mean opinion $\bar m$ and the single-community mean opinion $\bar m_{(1,1)}$ to see which steady state the system reaches. At the fully-mixed steady state, we expect that both $\bar m$ and $m_{(1,1)}$ are close to $0$. At a consensus steady state, neither $\bar m$ nor $\bar m_{(1,1)}$ is close to $0$. At a polarized steady state, we expect that $\bar m$ is close to $0$ but that $\bar m_{(1,1)}$ is not. For each set of parameters, we run our model on 10 networks. To calculate the mean opinions in our simulations, we first calculate the mean opinion of each node over the last 30 time steps. We then calculate the mean of these time-averaged opinions in the entire population and in community $C_{(1,1)}$. We take absolute values of these means because there is an equal probability for the population or a community to favor each opinion. Finally, we report the means of these absolute values (the population-scale and the community-scale time-averaged opinions) of the simulations. When we refer to ``mean opinions" from simulations in the rest of this section, we refer to the means that we calculate using this procedure. 


\subsection{Experiment 1}

{We set $f_2=0.5$ so that a node treats neighbors in the two layers equally when the majority opinions of the node's neighborhoods are different in the two layers.}
With this choice of $f_2$, our majority-vote model is the same as the model that was studied in Ref.~\cite{krawiecki2018majority} if one ignores the case that an equal number of neighbors hold each opinion in a layer.

\begin{figure*}
    \centering
    \includegraphics[scale=0.215]{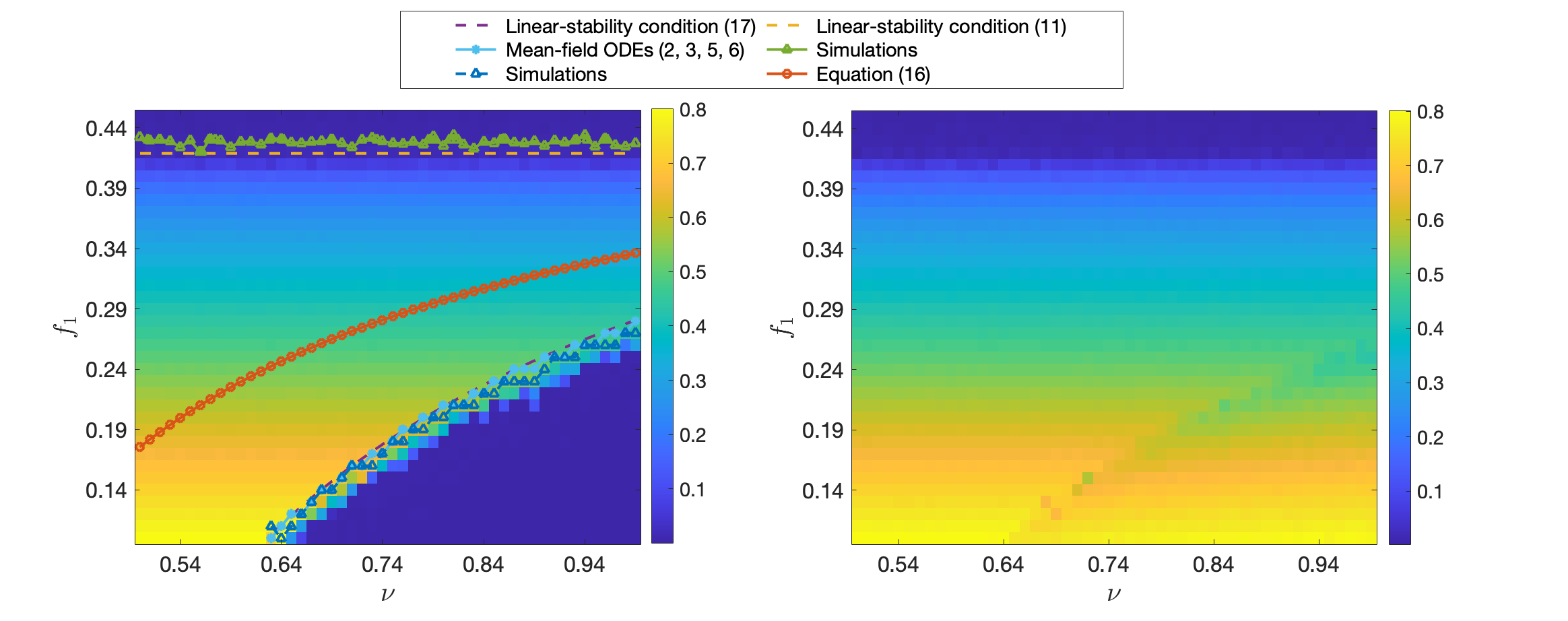}
    \caption{Heat maps of (left) the mean opinions of the entire population and (right) the mean opinions of community $C_{(1,1)}$ from means of $10$ direct numerical simulations. The horizontal axis is the parameter $\nu$ that controls the correlation of community assignments across layers and the vertical axis is the noise parameter $f_1$. In the left panel, we mark the boundaries of the stability regions of each steady state that we obtain from direct numerical simulations and from our mean-field approximation. We set $f_2 = 0.5$, $N = 15000$, and $\mu=0.25$.
    }
    \label{fig:mv-mf}
\end{figure*}

\begin{figure*}
    \centering
    \includegraphics[scale=0.215]{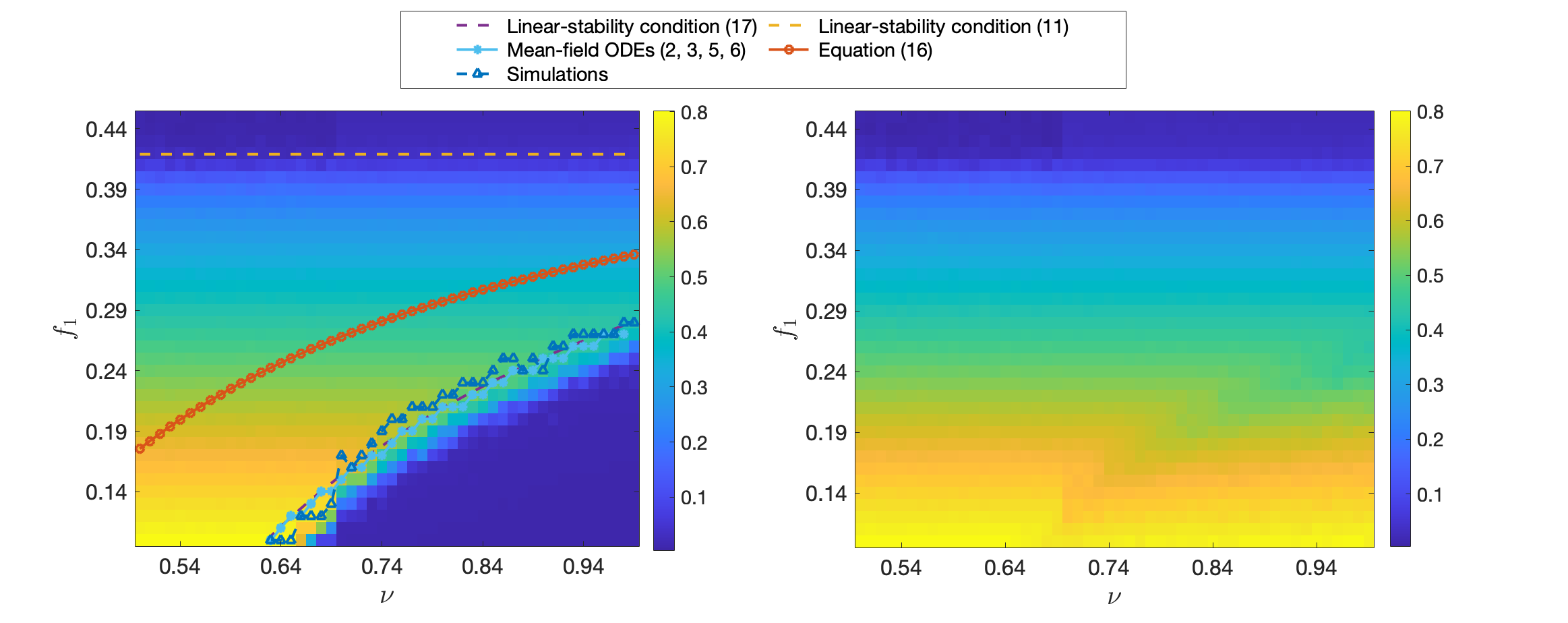}
    \caption{Heat maps of (left) the mean opinions of the entire population and (right) the mean opinions of community $C_{(1,1)}$ from means of $10$ direct numerical simulations. In the left panel, we mark the boundaries of the stability regions of each steady state that we obtain from direct simulations and from our mean-field approximation. We set $f_2 = 0.5$, $N = 5000$, and $\mu=0.25$.}
    \label{fig:mv-mf2}
\end{figure*}

In Figure \ref{fig:mv-mf}, we compare our simulation results and our mean-field results for different values of $f_1$ and $\nu$. We consider networks with $N=15000$ physical nodes. For each set of parameters, we sample $10$ networks from the SBM and run our model once on each network. We show the results of our simulations as heat maps. In this figure (and all subsequent similar figures), the left panel shows the mean opinions of the whole population and the right panel shows the mean opinions of community $C_{(1,1)}$. The heat maps in both panels have three regions. In the left panel, the middle region with values that are far away from $0$ separates the bottom-right and the top regions, which have values that are close to $0$. In the right panel, the top region has values that are close to $0$. The middle and the bottom-right regions both have values that are far away from $0$. The colors undergo sharper changes at locations where the two regions meet than inside the regions. The fully-mixed steady state is stable in the top region, and the consensus steady states are stable in the middle region. In the bottom-right region, both the consensus steady states and the polarized steady states are stable. Because of our choice of initial opinion distributions, the system reaches polarized steady states for almost every simulation except when the parameters are close to the boundary of the bottom-right region. Based on our numerical simulations, it seems that the mean opinions undergo a smooth transition between the fully-mixed region and the consensus region but that there is an abrupt transition between the consensus region and the polarized region. 

We mark the boundaries of the stability regions in the left panel of Figure \ref{fig:mv-mf}. For the transition between the fully-mixed regime and the consensus regime, we calculate an approximate $0$ level set of the mean opinions from our simulation results. Because we take absolute values when calculating the mean opinions, the values are larger than $0$. We use \textsc{Matlab}'s {\texttt contour} function \footnote{\textsc{Matlab}'s {\texttt contour} function finds isolines of a matrix through interpolation.} to find the parameter values at which the mean opinion equals $0.01$. (See the green curve with triangles.) We plot the linear-stability boundaries using \eqref{equ: mv-stability-first}. (See the yellow dashed curves.) These two curves are close to each other. For the transition between the consensus regime and the polarized regime, we observe a rapid decrease of the mean opinions when we decrease $f_1$, so we do not try to plot an approximate $0$ level set. Instead, we determine the values of $f_1$ that yield the maximum mean opinions for each $\nu$. (See the blue curve with triangles.) From our linear-stability condition \eqref{equ: mv-stability-2}, for each pair of values of $f$ and $\nu$, we solve for $\epsilon_1$ using fixed-point iterations and calculate the value of the largest eigenvalue. We collect the largest eigenvalues and obtain a matrix of eigenvalues. We calculate an approximate $0$ level set of this matrix using the {\texttt contour} function. (See the purple dashed curve.) We obtain a matrix of mean opinions by numerically solving the ODE system (\ref{equ: mv-eq2}, \ref{equ: mv-eq3}, \ref{equ: mv-eq5}, \ref{equ: mv-eq6}) for each pair of values of $f$ and $\nu$. The light-blue curve with asterisks indicates the approximate $0$ level set of this mean-opinion matrix. We set the initial condition to be compatible with the initial opinion distributions in our direct numerical simulations; this entails that $(\bar q_{(1,1)}, \bar q_{(1,2)},\bar q_{(2,1)},\bar q_{(2,2)}) = (1, 1,0, 0)$. The three curves are close to each other, but they are not very close to the approximate $0$ level set of the mean opinions from our simulation results. As the parameters approach the boundary of the stability region of the polarized steady states, it seems that the polarized steady states become less stable and the stochasticity of our numerical simulations makes it easier for the system to reach a consensus steady state. We also expect that the stochasticity has more influence for progressively smaller $N$ and that the gaps between the three curves and the actual $0$ level set decreases as $N$ increases. To examine this, we run the same simulations for $N=5000$ and plot our results in Figure \ref{fig:mv-mf2}. The gaps widen, as expected. Finally, we plot the condition \eqref{equ: mv-eq8} (see the red curve with circles) to verify that condition \eqref{equ: mv-eq8} is a necessary but not sufficient condition for polarized steady states to be linearly stable. 

Overall, we observe that the phase-transition patterns for $f_2 = 0.5$ are qualitatively the same as what was obtained for the monolayer majority-vote model in Ref.~\cite{huang2015phase}. For a fixed $\nu$, the system undergoes two transitions as $f_1$ decreases. The smaller critical value is independent of $\nu$ (as we see in our mean-field 
description) and the larger critical value increases with $\nu$.


\subsection{Experiment 2}

\begin{figure*}
    \centering
     \includegraphics[scale=0.2]{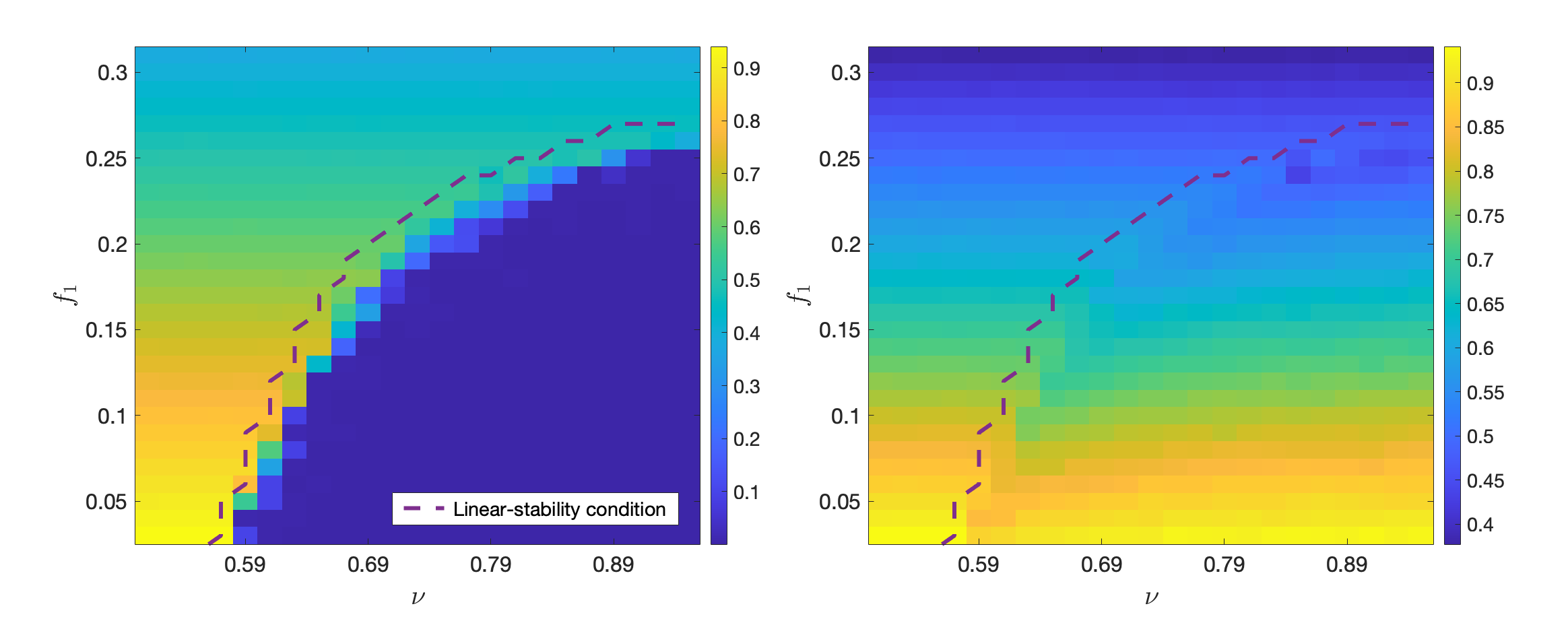}
     \caption{Heat maps of (left) the mean opinions of the entire population and (right) the mean opinions of community $C_{(1,1)}$ from means of $10$ direct numerical simulations. In both panels, we mark the boundary of the linear-stability region of polarized steady states; we obtain it by setting $ \lambda_{\text{polarized}} = 0$ in equation \eqref{equ: mv-polarization_condition}. We set $f_2 = 0.32$, $N = 10000$, and $\mu=0.25$.}\label{fig:mv-mf2-2}
\end{figure*}

\begin{figure*}
    \centering
     \includegraphics[scale=0.2]{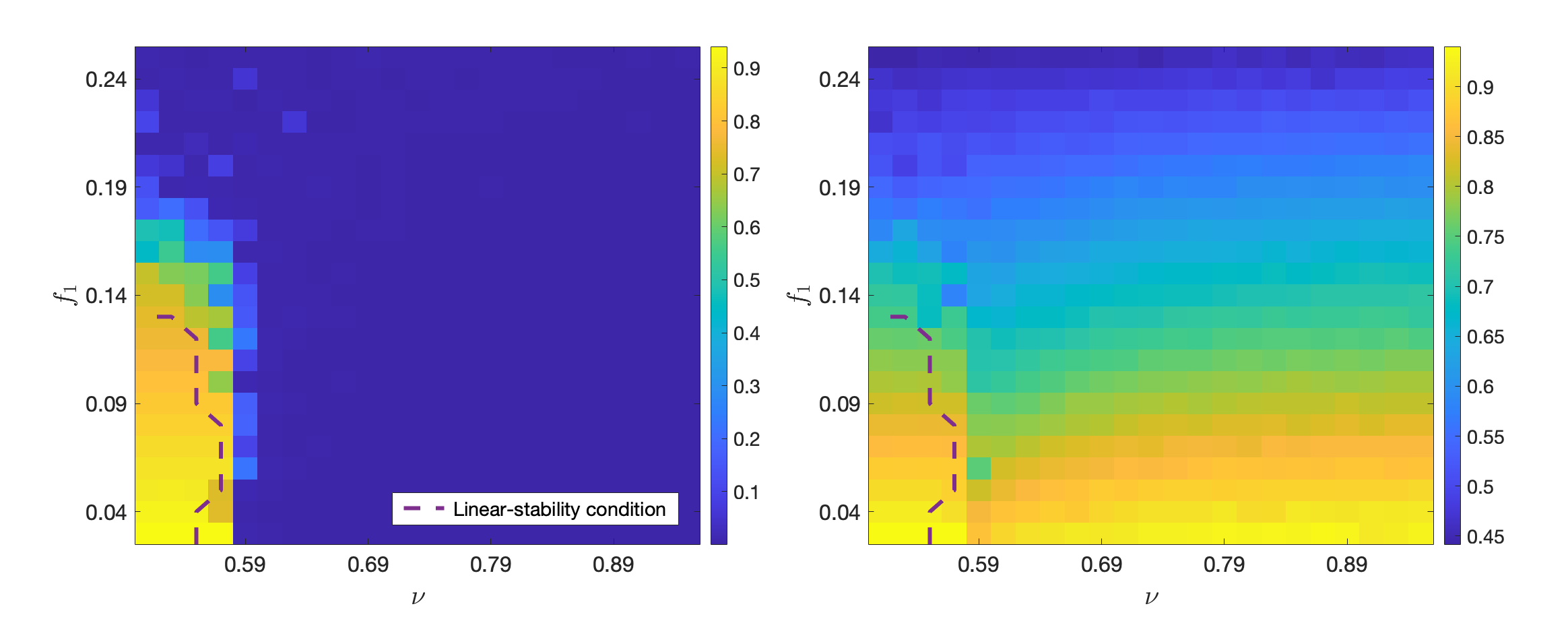}
     \caption{Heat maps of (left) the mean opinions of the entire population and (right) the mean opinions of community $C_{(1,1)}$ from means of $10$ direct numerical simulations. In both panels, we mark the boundary of the linear-stability region of polarized steady states; we obtain it by setting $ \lambda_{\text{polarized}} = 0$ in equation \eqref{equ: mv-polarization_condition}. We set $f_2 = 0.26$, $N = 10000$, and $\mu=0.25$.}\label{fig:mv-mf2-1}
\end{figure*}

In this experiment, we explore the influence of the layer-preference parameter $f_2$ on the steady-state results. In Figures \ref{fig:mv-mf2-2} and \ref{fig:mv-mf2-1}, we show heat maps of the mean opinions for a range of $f_1$ and $\nu$ values when $f_2 = 0.32$ and $f_2 = 0.26$, respectively. These two figures show two typical heat maps. In the left panels of both figures, the dark-blue regions indicate regimes in which both consensus steady states and polarized steady states are stable; in the yellow regions, only consensus steady states are stable. At locations where the two regions meet, both figures have jagged boundaries, which one can smoothen by running simulations with more finely-grained values of $f_1$ and $\nu$. Figure \ref{fig:mv-mf2-2} has a similar pattern as Figure \ref{fig:mv-mf}. In Figure \ref{fig:mv-mf2-1}, however, the polarized steady states are stable in a much larger region than in Figures \ref{fig:mv-mf} and \ref{fig:mv-mf2-2}. This indicates that it is easier for different communities to favor different opinions when the nodes have a stronger preference for opinions in one layer than for those in the other. Additionally, when $\nu$ is small, polarized steady states are stable for large values of $f_1$ but not for small values of $f_1$. We do not know the reason that small values of $f_1$ appear to lead to unstable polarized steady states.

\subsection{Experiment 3}

\begin{figure*}
    \centering
     \includegraphics[scale=0.2]{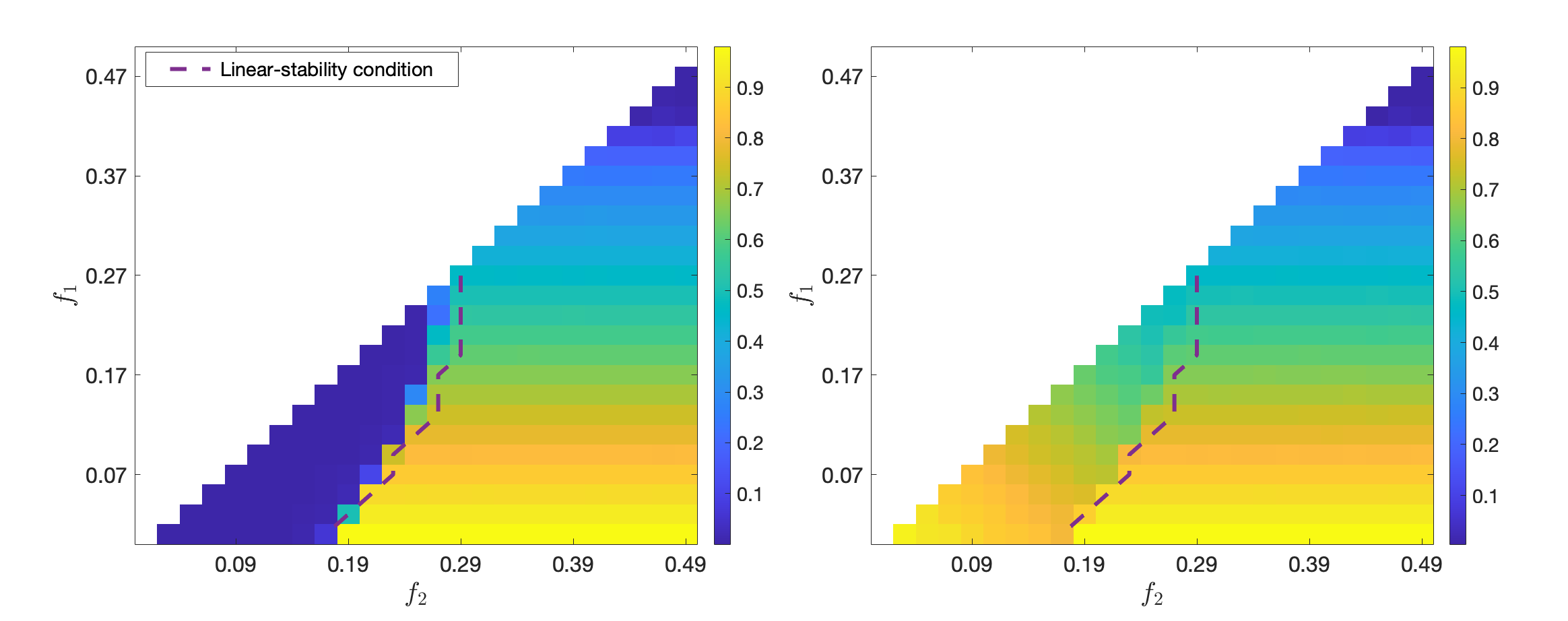}
     \caption{Heat maps of (left) the mean opinions of the entire population and (right) the mean opinions of community $C_{(1,1)}$ from means of $10$ direct numerical simulations. In both panels, we mark the boundary of the linear-stability region of polarized steady states; we obtain it by setting $ \lambda_{\text{polarized}} = 0$ in equation \eqref{equ: mv-polarization_condition}. We set $\nu = 0.5$, $N = 10000$, and $\mu=0.25$.}    \label{fig:mv-mf3}\label{fig:mv-mf3-1}
\end{figure*}

\begin{figure*}
    \centering
     \includegraphics[scale=0.2, trim = 0 0 0 0, clip]{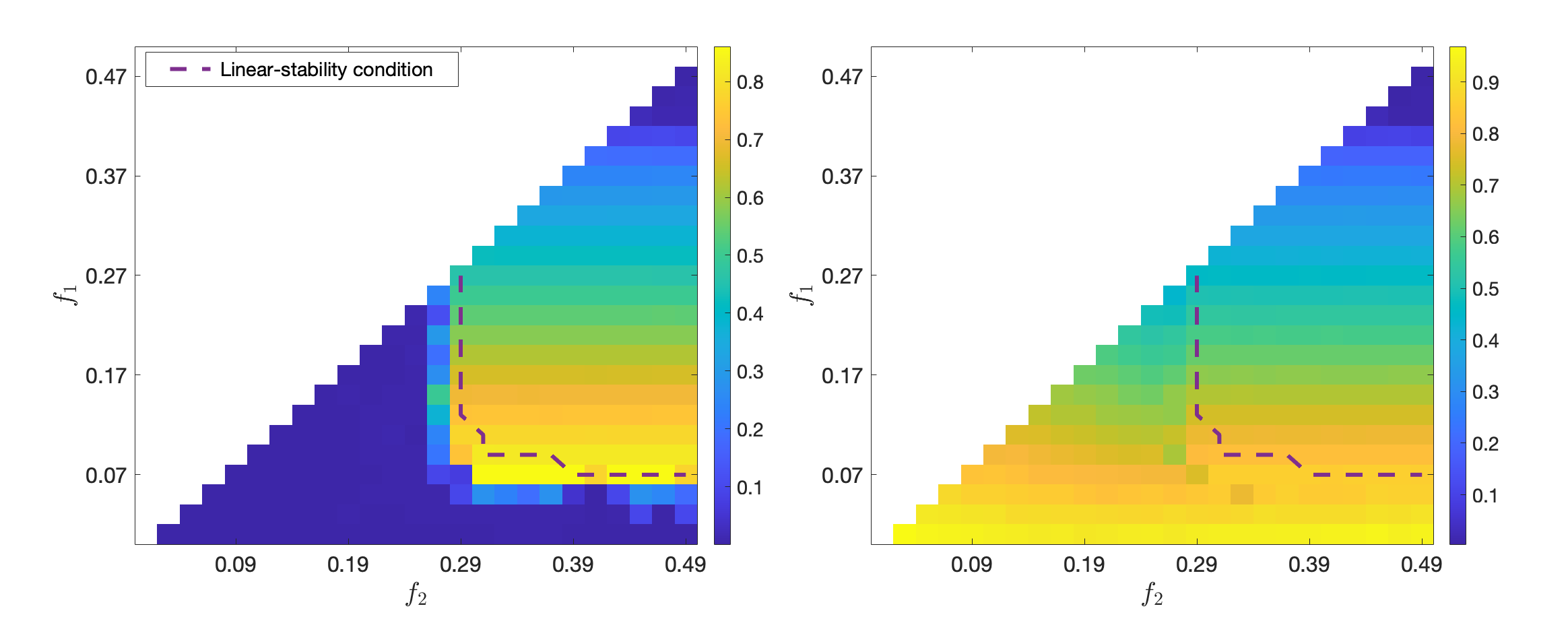}
     \caption{Heat maps of (left) the mean opinions of the entire population and (right) the mean opinions of community $C_{(1,1)}$ from means of $10$ direct numerical simulations. In both panels, we mark the boundary of the linear-stability region of polarized steady states; we obtain it by setting $ \lambda_{\text{polarized}} = 0$ in equation \eqref{equ: mv-polarization_condition}. We set $\nu = 0.6$, $N = 10000$, and $\mu=0.25$.}\label{fig:mv-mf3-2}
\end{figure*}

\begin{figure*}
 \includegraphics[scale=0.2, trim = 0 0 0 0, clip]{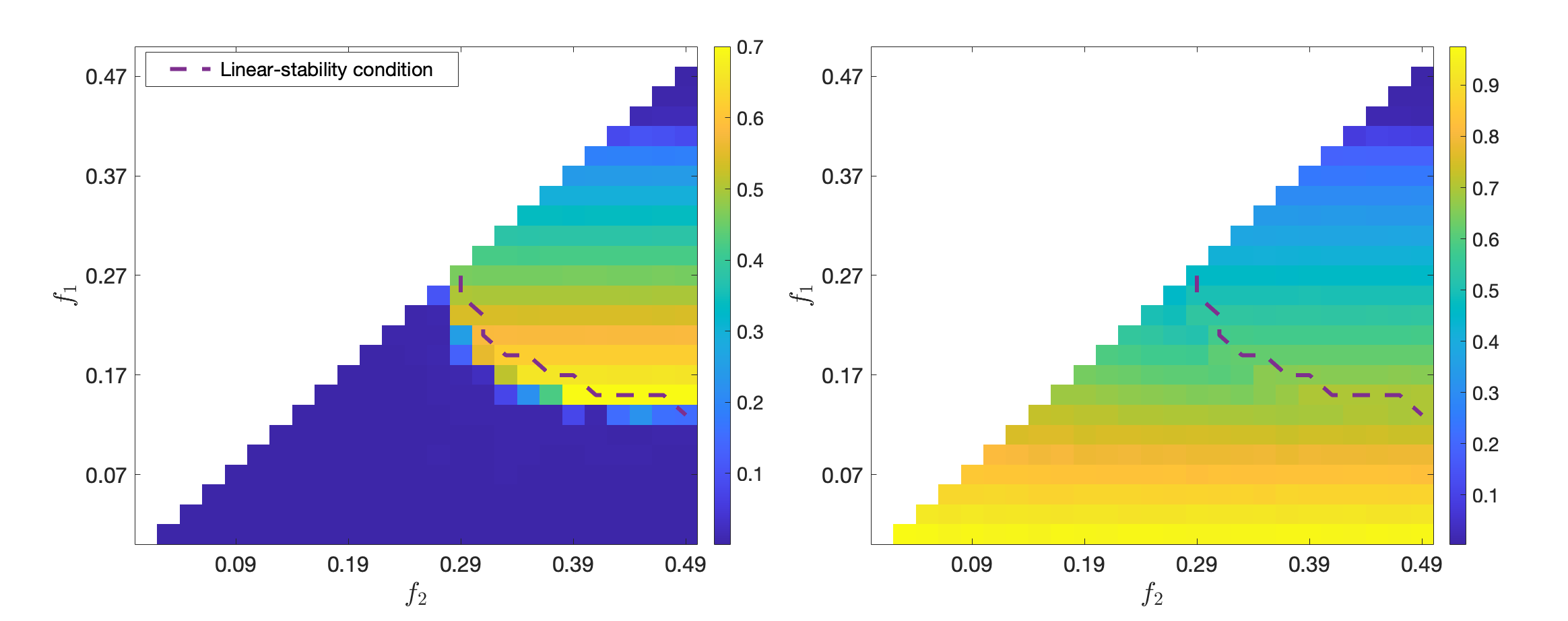}
 \caption{Heat maps of (left) the mean opinions of the entire population and (right) the mean opinions of community $C_{(1,1)}$ from means of $10$ direct numerical simulations. In both panels, we mark the boundary of the linear-stability region of polarized steady states; we obtain it by setting $ \lambda_{\text{polarized}} = 0$ in equation \eqref{equ: mv-polarization_condition}. We set $\nu = 0.7$, $N = 10000$, and $\mu=0.25$.}\label{fig:mv-mf3-3}
\end{figure*}

In this experiment, we fix the interlayer community correlation $\nu$ and examine our model for different values of $f_1$ and $f_2$ with $0.5\geq f_2>f_1>0$. We show heat maps of the mean opinions in Figure \ref{fig:mv-mf3-1}--\ref{fig:mv-mf3-3} for different values of $\nu$. The dark-blue regions in the lower part of the triangles are regions where polarized steady states are stable. From our linear stability analysis, we obtain the boundaries of linear-stability regions of polarized steady states by setting $ \lambda_{\text{polarized}} = 0$ in equation \eqref{equ: mv-polarization_condition}. We mark the boundaries with dashed curves. The area of this region increases as we increase $\nu$, so it is easier for different communities in a layer to develop preferences for different opinions when the communities in the two layers are more strongly correlated with each other. We also observe some other interesting features. Let $f_{1}^*$ denote the maximum value of $f_1$ for which there exists some $f_2$ such that polarized steady states are stable. When $f_1>f_{1}^*$, polarized steady states are not stable for any value of $f_2$. The value of $f_{1}^*$ is the same in Figures \ref{fig:mv-mf3-1}--\ref{fig:mv-mf3-3}. Let $f_2^*$ denote the value of $f_2$ such that $(f_1^*, f_2^*)$ is on the boundary of the stability region. When $\nu$ is small (see Figure \ref{fig:mv-mf3-1}), there exists $f_2 < f_2^*$ such that polarized steady states are stable for some $f_1$. Otherwise, $f_2^*$ is the minimum value of $f_2$ such that polarized steady states are stable. (See Figures \ref{fig:mv-mf3-2} and \ref{fig:mv-mf3-3}.)

\section{Conclusions and discussion} \label{sec: mv-conclusion}

We proposed a multiplex majority-vote model in which each layer represents one relationship and each physical node holds one opinion. Our model considers situations in which different relationships have different propensities to influence an individual's opinion. This feature cannot exist in monolayer opinion models and (to the best of our knowledge) was neglected in the multiplex majority-vote models that were studied in previous works.

We examined the steady-state opinion distributions of our majority-vote model on networks with community structure with both a mean-field approximation and direct numerical simulations. We observed the same three regimes --- fully-mixed, consensus, and polarized steady states --- as those that have been observed in majority-vote models on monolayer networks with community structure \cite{huang2015phase}. When neighbors in different layers are equally influential (i.e., $f_2 = 0.5$), we found that the phase-transition patterns are qualitatively the same as what was obtained in Ref.~\cite{huang2015phase}. For a fixed interlayer community correlation $\nu$, the system undergoes two transitions as we decrease $f_1$. The smaller critical value is independent of $\nu$, but the larger critical value increases with $\nu$. When $f_2 \neq 0.5$, we showed that the heterogeneous influence abilities of different layers can qualitatively affect the regions in which polarized steady states are stable. We also found that a stronger interlayer community correlation results in polarized steady states being reachable for wider ranges of the parameter $f_1$ and $f_2$.

There are many ways to generalize our model. First, one can consider more complicated network structures. For example, one can examine more realistic community structures (e.g., by considering degree-corrected SBMs or overlapping communities), and one can also study the influence of interlayer degree correlations and interlayer edge overlaps on opinion spread. Second, one can consider more general opinion-update rules. For example, when the two layers have the same majority opinion, the probability that a node flips its opinion when the common majority opinion is the same as its opinion and the probability that it flips when the common majority opinion is different from its opinion do not have to sum to $1$. Third, in our model, each node has the same propensity of conforming to one of the two layers when the two layers have different majority opinions (i.e., they have the same value of $f_2$). A natural extension is to consider a model in which the nodes have heterogeneous propensities. This is typically relevant when the different layers represent different online social platforms; different people often do prefer different social platforms in real life. Moreover, because individuals who belong to the same communities often share common traits, it seems sensible to study situations in which individuals' preferred relationships are correlated with their community memberships.


\begin{acknowledgments}

This research is supported by NSF grants DMS-1922952 and DMS-1737770 through the Algorithms for Threat Detection (ATD) program.

\end{acknowledgments}



\begin{thebibliography}{51}%
\makeatletter
\providecommand \@ifxundefined [1]{%
 \@ifx{#1\undefined}
}%
\providecommand \@ifnum [1]{%
 \ifnum #1\expandafter \@firstoftwo
 \else \expandafter \@secondoftwo
 \fi
}%
\providecommand \@ifx [1]{%
 \ifx #1\expandafter \@firstoftwo
 \else \expandafter \@secondoftwo
 \fi
}%
\providecommand \natexlab [1]{#1}%
\providecommand \enquote  [1]{``#1''}%
\providecommand \bibnamefont  [1]{#1}%
\providecommand \bibfnamefont [1]{#1}%
\providecommand \citenamefont [1]{#1}%
\providecommand \href@noop [0]{\@secondoftwo}%
\providecommand \href [0]{\begingroup \@sanitize@url \@href}%
\providecommand \@href[1]{\@@startlink{#1}\@@href}%
\providecommand \@@href[1]{\endgroup#1\@@endlink}%
\providecommand \@sanitize@url [0]{\catcode `\\12\catcode `\$12\catcode
  `\&12\catcode `\#12\catcode `\^12\catcode `\_12\catcode `\%12\relax}%
\providecommand \@@startlink[1]{}%
\providecommand \@@endlink[0]{}%
\providecommand \url  [0]{\begingroup\@sanitize@url \@url }%
\providecommand \@url [1]{\endgroup\@href {#1}{\urlprefix }}%
\providecommand \urlprefix  [0]{URL }%
\providecommand \Eprint [0]{\href }%
\providecommand \doibase [0]{https://doi.org/}%
\providecommand \selectlanguage [0]{\@gobble}%
\providecommand \bibinfo  [0]{\@secondoftwo}%
\providecommand \bibfield  [0]{\@secondoftwo}%
\providecommand \translation [1]{[#1]}%
\providecommand \BibitemOpen [0]{}%
\providecommand \bibitemStop [0]{}%
\providecommand \bibitemNoStop [0]{.\EOS\space}%
\providecommand \EOS [0]{\spacefactor3000\relax}%
\providecommand \BibitemShut  [1]{\csname bibitem#1\endcsname}%
\let\auto@bib@innerbib\@empty
\bibitem [{\citenamefont {Asch}(1951)}]{asch1951effects}%
  \BibitemOpen
  \bibfield  {author} {\bibinfo {author} {\bibfnamefont {S.~E.}\ \bibnamefont
  {Asch}},\ }\bibfield  {title} {\bibinfo {title} {Effects of group pressure
  upon the modification and distortion of judgments},\ }in\ \href@noop {}
  {\emph {\bibinfo {booktitle} {Groups, Leadership and Men: Research in Human
  Relations}}},\ \bibinfo {editor} {edited by\ \bibinfo {editor} {\bibfnamefont
  {H.}~\bibnamefont {Guetzkow}}}\ (\bibinfo  {publisher} {Carnegie Press},\
  \bibinfo {address} {Pittsburgh, PA},\ \bibinfo {year} {1951})\ pp.\ \bibinfo
  {pages} {177--190}\BibitemShut {NoStop}%
\bibitem [{\citenamefont {Flache}\ \emph {et~al.}(2017)\citenamefont {Flache},
  \citenamefont {M\"{a}s}, \citenamefont {Feliciani}, \citenamefont
  {Chattoe-Brown}, \citenamefont {Deffuant}, \citenamefont {Huet},\ and\
  \citenamefont {Lorenz}}]{flache2017}%
  \BibitemOpen
  \bibfield  {author} {\bibinfo {author} {\bibfnamefont {A.}~\bibnamefont
  {Flache}}, \bibinfo {author} {\bibfnamefont {M.}~\bibnamefont {M\"{a}s}},
  \bibinfo {author} {\bibfnamefont {T.}~\bibnamefont {Feliciani}}, \bibinfo
  {author} {\bibfnamefont {E.}~\bibnamefont {Chattoe-Brown}}, \bibinfo {author}
  {\bibfnamefont {G.}~\bibnamefont {Deffuant}}, \bibinfo {author}
  {\bibfnamefont {S.}~\bibnamefont {Huet}},\ and\ \bibinfo {author}
  {\bibfnamefont {J.}~\bibnamefont {Lorenz}},\ }\bibfield  {title} {\bibinfo
  {title} {Models of social influence: {T}owards the next frontiers},\ }\href
  {https://doi.org/10.18564/jasss.3521} {\bibfield  {journal} {\bibinfo
  {journal} {Journal of Artificial Societies and Social Simulation}\ }\textbf
  {\bibinfo {volume} {20}},\ \bibinfo {pages} {2} (\bibinfo {year}
  {2017})}\BibitemShut {NoStop}%
\bibitem [{\citenamefont {Porter}\ and\ \citenamefont
  {Gleeson}(2016)}]{porter2016}%
  \BibitemOpen
  \bibfield  {author} {\bibinfo {author} {\bibfnamefont {M.~A.}\ \bibnamefont
  {Porter}}\ and\ \bibinfo {author} {\bibfnamefont {J.~P.}\ \bibnamefont
  {Gleeson}},\ }\href@noop {} {\emph {\bibinfo {title} {Dynamical Systems on
  Networks: {A} Tutorial}}},\ Vol.~\bibinfo {volume} {4}\ (\bibinfo
  {publisher} {Springer International Publishing},\ \bibinfo {address} {Cham,
  Switzerland},\ \bibinfo {year} {2016})\BibitemShut {NoStop}%
\bibitem [{\citenamefont {Newman}(2018)}]{newman2018networks}%
  \BibitemOpen
  \bibfield  {author} {\bibinfo {author} {\bibfnamefont {M.~E.~J.}\
  \bibnamefont {Newman}},\ }\href@noop {} {\emph {\bibinfo {title}
  {Networks}}},\ \bibinfo {edition} {2nd}\ ed.\ (\bibinfo  {publisher} {Oxford
  University Press},\ \bibinfo {address} {Oxford, UK},\ \bibinfo {year}
  {2018})\BibitemShut {NoStop}%
\bibitem [{\citenamefont {Castellano}\ \emph {et~al.}(2009)\citenamefont
  {Castellano}, \citenamefont {Fortunato},\ and\ \citenamefont
  {Loreto}}]{RevModPhys.81.591}%
  \BibitemOpen
  \bibfield  {author} {\bibinfo {author} {\bibfnamefont {C.}~\bibnamefont
  {Castellano}}, \bibinfo {author} {\bibfnamefont {S.}~\bibnamefont
  {Fortunato}},\ and\ \bibinfo {author} {\bibfnamefont {V.}~\bibnamefont
  {Loreto}},\ }\bibfield  {title} {\bibinfo {title} {Statistical physics of
  social dynamics},\ }\href {https://doi.org/10.1103/RevModPhys.81.591}
  {\bibfield  {journal} {\bibinfo  {journal} {Reviews of Modern Physics}\
  }\textbf {\bibinfo {volume} {81}},\ \bibinfo {pages} {591} (\bibinfo {year}
  {2009})}\BibitemShut {NoStop}%
\bibitem [{\citenamefont {Noorazar}\ \emph {et~al.}(2020)\citenamefont
  {Noorazar}, \citenamefont {Vixie}, \citenamefont {Talebanpour},\ and\
  \citenamefont {Hu}}]{noorazar2020classical}%
  \BibitemOpen
  \bibfield  {author} {\bibinfo {author} {\bibfnamefont {H.}~\bibnamefont
  {Noorazar}}, \bibinfo {author} {\bibfnamefont {K.~R.}\ \bibnamefont {Vixie}},
  \bibinfo {author} {\bibfnamefont {A.}~\bibnamefont {Talebanpour}},\ and\
  \bibinfo {author} {\bibfnamefont {Y.}~\bibnamefont {Hu}},\ }\bibfield
  {title} {\bibinfo {title} {From classical to modern opinion dynamics},\
  }\href@noop {} {\bibfield  {journal} {\bibinfo  {journal} {International
  Journal of Modern Physics C}\ }\textbf {\bibinfo {volume} {31}},\ \bibinfo
  {pages} {2050101} (\bibinfo {year} {2020})}\BibitemShut {NoStop}%
\bibitem [{\citenamefont {Peralta}\ \emph {et~al.}(2022)\citenamefont
  {Peralta}, \citenamefont {Kert{\'e}sz},\ and\ \citenamefont
  {I{\~n}iguez}}]{peralta2022opinion}%
  \BibitemOpen
  \bibfield  {author} {\bibinfo {author} {\bibfnamefont {A.~F.}\ \bibnamefont
  {Peralta}}, \bibinfo {author} {\bibfnamefont {J.}~\bibnamefont
  {Kert{\'e}sz}},\ and\ \bibinfo {author} {\bibfnamefont {G.}~\bibnamefont
  {I{\~n}iguez}},\ }\bibfield  {title} {\bibinfo {title} {Opinion dynamics in
  social networks: From models to data},\ }\href@noop {} {\bibfield  {journal}
  {\bibinfo  {journal} {arXiv preprint arXiv:2201.01322}\ } (\bibinfo {year}
  {2022})}\BibitemShut {NoStop}%
\bibitem [{\citenamefont {de~Oliveira}(1992)}]{de1992isotropic}%
  \BibitemOpen
  \bibfield  {author} {\bibinfo {author} {\bibfnamefont {M.~J.}\ \bibnamefont
  {de~Oliveira}},\ }\bibfield  {title} {\bibinfo {title} {Isotropic
  majority-vote model on a square lattice},\ }\href@noop {} {\bibfield
  {journal} {\bibinfo  {journal} {Journal of Statistical Physics}\ }\textbf
  {\bibinfo {volume} {66}},\ \bibinfo {pages} {273} (\bibinfo {year}
  {1992})}\BibitemShut {NoStop}%
\bibitem [{\citenamefont {Vilela}\ \emph {et~al.}(2019)\citenamefont {Vilela},
  \citenamefont {Wang}, \citenamefont {Nelson},\ and\ \citenamefont
  {Stanley}}]{vilela2019majority}%
  \BibitemOpen
  \bibfield  {author} {\bibinfo {author} {\bibfnamefont {A.~L.}\ \bibnamefont
  {Vilela}}, \bibinfo {author} {\bibfnamefont {C.}~\bibnamefont {Wang}},
  \bibinfo {author} {\bibfnamefont {K.~P.}\ \bibnamefont {Nelson}},\ and\
  \bibinfo {author} {\bibfnamefont {H.~E.}\ \bibnamefont {Stanley}},\
  }\bibfield  {title} {\bibinfo {title} {Majority-vote model for financial
  markets},\ }\href@noop {} {\bibfield  {journal} {\bibinfo  {journal} {Physica
  A: Statistical Mechanics and its Applications}\ }\textbf {\bibinfo {volume}
  {515}},\ \bibinfo {pages} {762} (\bibinfo {year} {2019})}\BibitemShut
  {NoStop}%
\bibitem [{\citenamefont {Lima}(2010)}]{lima2010analysing}%
  \BibitemOpen
  \bibfield  {author} {\bibinfo {author} {\bibfnamefont {F.~W.~S.}\
  \bibnamefont {Lima}},\ }\bibfield  {title} {\bibinfo {title} {Analysing and
  controlling the tax evasion dynamics via majority-vote model},\ }\href@noop
  {} {\bibfield  {journal} {\bibinfo  {journal} {Journal of Physics: Conference
  Series}\ }\textbf {\bibinfo {volume} {246}},\ \bibinfo {pages} {012033}
  (\bibinfo {year} {2010})}\BibitemShut {NoStop}%
\bibitem [{\citenamefont {Vilela}\ \emph {et~al.}(2021)\citenamefont {Vilela},
  \citenamefont {Pereira}, \citenamefont {Dias}, \citenamefont {Stanley},\ and\
  \citenamefont {da~Silva}}]{vilela2021majority}%
  \BibitemOpen
  \bibfield  {author} {\bibinfo {author} {\bibfnamefont {A.~L.~M.}\
  \bibnamefont {Vilela}}, \bibinfo {author} {\bibfnamefont {L.~F.~C.}\
  \bibnamefont {Pereira}}, \bibinfo {author} {\bibfnamefont {L.}~\bibnamefont
  {Dias}}, \bibinfo {author} {\bibfnamefont {H.~E.}\ \bibnamefont {Stanley}},\
  and\ \bibinfo {author} {\bibfnamefont {L.~R.}\ \bibnamefont {da~Silva}},\
  }\bibfield  {title} {\bibinfo {title} {Majority-vote model with limited
  visibility: {A}n investigation into filter bubbles},\ }\href@noop {}
  {\bibfield  {journal} {\bibinfo  {journal} {Physica A: Statistical Mechanics
  and its Applications}\ }\textbf {\bibinfo {volume} {563}},\ \bibinfo {pages}
  {125450} (\bibinfo {year} {2021})}\BibitemShut {NoStop}%
\bibitem [{\citenamefont {Pereira}\ and\ \citenamefont
  {Moreira}(2005)}]{pereira2005majority}%
  \BibitemOpen
  \bibfield  {author} {\bibinfo {author} {\bibfnamefont {L.~F.}\ \bibnamefont
  {Pereira}}\ and\ \bibinfo {author} {\bibfnamefont {F.~B.}\ \bibnamefont
  {Moreira}},\ }\bibfield  {title} {\bibinfo {title} {Majority-vote model on
  random graphs},\ }\href@noop {} {\bibfield  {journal} {\bibinfo  {journal}
  {Physical Review E}\ }\textbf {\bibinfo {volume} {71}},\ \bibinfo {pages}
  {016123} (\bibinfo {year} {2005})}\BibitemShut {NoStop}%
\bibitem [{\citenamefont {Lima}\ \emph {et~al.}(2008)\citenamefont {Lima},
  \citenamefont {Sousa},\ and\ \citenamefont {Sumuor}}]{lima2008majority}%
  \BibitemOpen
  \bibfield  {author} {\bibinfo {author} {\bibfnamefont {F.~W.~S.}\
  \bibnamefont {Lima}}, \bibinfo {author} {\bibfnamefont {A.~O.}\ \bibnamefont
  {Sousa}},\ and\ \bibinfo {author} {\bibfnamefont {M.~A.}\ \bibnamefont
  {Sumuor}},\ }\bibfield  {title} {\bibinfo {title} {Majority-vote on directed
  {Erd{\H{o}}s--R{\'e}nyi} random graphs},\ }\href@noop {} {\bibfield
  {journal} {\bibinfo  {journal} {Physica A: Statistical Mechanics and its
  Applications}\ }\textbf {\bibinfo {volume} {387}},\ \bibinfo {pages} {3503}
  (\bibinfo {year} {2008})}\BibitemShut {NoStop}%
\bibitem [{\citenamefont {Campos}\ \emph {et~al.}(2003)\citenamefont {Campos},
  \citenamefont {de~Oliveira},\ and\ \citenamefont
  {Moreira}}]{campos2003small}%
  \BibitemOpen
  \bibfield  {author} {\bibinfo {author} {\bibfnamefont {P.~R.~A.}\
  \bibnamefont {Campos}}, \bibinfo {author} {\bibfnamefont {V.~M.}\
  \bibnamefont {de~Oliveira}},\ and\ \bibinfo {author} {\bibfnamefont
  {F.~G.~B.}\ \bibnamefont {Moreira}},\ }\bibfield  {title} {\bibinfo {title}
  {Small-world effects in the majority-vote model},\ }\href@noop {} {\bibfield
  {journal} {\bibinfo  {journal} {Physical Review E}\ }\textbf {\bibinfo
  {volume} {67}},\ \bibinfo {pages} {026104} (\bibinfo {year}
  {2003})}\BibitemShut {NoStop}%
\bibitem [{\citenamefont {Luz}\ and\ \citenamefont
  {Lima}(2007)}]{luz2007majority}%
  \BibitemOpen
  \bibfield  {author} {\bibinfo {author} {\bibfnamefont {E.~M.~S.}\
  \bibnamefont {Luz}}\ and\ \bibinfo {author} {\bibfnamefont {F.~W.~S.}\
  \bibnamefont {Lima}},\ }\bibfield  {title} {\bibinfo {title} {Majority-vote
  on directed small-world networks},\ }\href@noop {} {\bibfield  {journal}
  {\bibinfo  {journal} {International Journal of Modern Physics C}\ }\textbf
  {\bibinfo {volume} {18}},\ \bibinfo {pages} {1251} (\bibinfo {year}
  {2007})}\BibitemShut {NoStop}%
\bibitem [{\citenamefont {Stone}\ and\ \citenamefont
  {McKay}(2015)}]{stone2015majority}%
  \BibitemOpen
  \bibfield  {author} {\bibinfo {author} {\bibfnamefont {T.~E.}\ \bibnamefont
  {Stone}}\ and\ \bibinfo {author} {\bibfnamefont {S.~R.}\ \bibnamefont
  {McKay}},\ }\bibfield  {title} {\bibinfo {title} {Majority-vote model on a
  dynamic small-world network},\ }\href@noop {} {\bibfield  {journal} {\bibinfo
   {journal} {Physica A: Statistical Mechanics and its Applications}\ }\textbf
  {\bibinfo {volume} {419}},\ \bibinfo {pages} {437} (\bibinfo {year}
  {2015})}\BibitemShut {NoStop}%
\bibitem [{\citenamefont {Lima}(2006)}]{lima2006majority}%
  \BibitemOpen
  \bibfield  {author} {\bibinfo {author} {\bibfnamefont {F.~W.~S.}\
  \bibnamefont {Lima}},\ }\bibfield  {title} {\bibinfo {title} {Majority-vote
  on directed {Barab\'{a}si--Albert} networks},\ }\href@noop {} {\bibfield
  {journal} {\bibinfo  {journal} {International Journal of Modern Physics C}\
  }\textbf {\bibinfo {volume} {17}},\ \bibinfo {pages} {1257} (\bibinfo {year}
  {2006})}\BibitemShut {NoStop}%
\bibitem [{\citenamefont {Huang}\ \emph {et~al.}(2015)\citenamefont {Huang},
  \citenamefont {Chen},\ and\ \citenamefont {Shen}}]{huang2015phase}%
  \BibitemOpen
  \bibfield  {author} {\bibinfo {author} {\bibfnamefont {F.}~\bibnamefont
  {Huang}}, \bibinfo {author} {\bibfnamefont {H.-S.}\ \bibnamefont {Chen}},\
  and\ \bibinfo {author} {\bibfnamefont {C.-S.}\ \bibnamefont {Shen}},\
  }\bibfield  {title} {\bibinfo {title} {Phase transitions of majority-vote
  model on modular networks},\ }\href@noop {} {\bibfield  {journal} {\bibinfo
  {journal} {Chinese Physics Letters}\ }\textbf {\bibinfo {volume} {32}},\
  \bibinfo {pages} {118902} (\bibinfo {year} {2015})}\BibitemShut {NoStop}%
\bibitem [{\citenamefont {Yang}\ \emph {et~al.}(2008)\citenamefont {Yang},
  \citenamefont {Kim},\ and\ \citenamefont {Kwak}}]{yang2008existence}%
  \BibitemOpen
  \bibfield  {author} {\bibinfo {author} {\bibfnamefont {J.-S.}\ \bibnamefont
  {Yang}}, \bibinfo {author} {\bibfnamefont {I.-m.}\ \bibnamefont {Kim}},\ and\
  \bibinfo {author} {\bibfnamefont {W.}~\bibnamefont {Kwak}},\ }\bibfield
  {title} {\bibinfo {title} {Existence of an upper critical dimension in the
  majority voter model},\ }\href@noop {} {\bibfield  {journal} {\bibinfo
  {journal} {Physical Review E}\ }\textbf {\bibinfo {volume} {77}},\ \bibinfo
  {pages} {051122} (\bibinfo {year} {2008})}\BibitemShut {NoStop}%
\bibitem [{\citenamefont {Wu}\ and\ \citenamefont
  {Holme}(2010)}]{wu2010majority}%
  \BibitemOpen
  \bibfield  {author} {\bibinfo {author} {\bibfnamefont {Z.-X.}\ \bibnamefont
  {Wu}}\ and\ \bibinfo {author} {\bibfnamefont {P.}~\bibnamefont {Holme}},\
  }\bibfield  {title} {\bibinfo {title} {Majority-vote model on hyperbolic
  lattices},\ }\href@noop {} {\bibfield  {journal} {\bibinfo  {journal}
  {Physical Review E}\ }\textbf {\bibinfo {volume} {81}},\ \bibinfo {pages}
  {011133} (\bibinfo {year} {2010})}\BibitemShut {NoStop}%
\bibitem [{\citenamefont {Santos}\ \emph {et~al.}(2011)\citenamefont {Santos},
  \citenamefont {Lima},\ and\ \citenamefont {Malarz}}]{santos2011majority}%
  \BibitemOpen
  \bibfield  {author} {\bibinfo {author} {\bibfnamefont {J.~C.}\ \bibnamefont
  {Santos}}, \bibinfo {author} {\bibfnamefont {F.~W.~S.}\ \bibnamefont
  {Lima}},\ and\ \bibinfo {author} {\bibfnamefont {K.}~\bibnamefont {Malarz}},\
  }\bibfield  {title} {\bibinfo {title} {Majority-vote model on triangular,
  honeycomb and kagom{\'e} lattices},\ }\href@noop {} {\bibfield  {journal}
  {\bibinfo  {journal} {Physica A: Statistical Mechanics and its Applications}\
  }\textbf {\bibinfo {volume} {390}},\ \bibinfo {pages} {359} (\bibinfo {year}
  {2011})}\BibitemShut {NoStop}%
\bibitem [{\citenamefont {Acu{\~n}a-Lara}\ \emph {et~al.}(2014)\citenamefont
  {Acu{\~n}a-Lara}, \citenamefont {Sastre},\ and\ \citenamefont
  {Vargas-Arriola}}]{acuna2014critical}%
  \BibitemOpen
  \bibfield  {author} {\bibinfo {author} {\bibfnamefont {A.~L.}\ \bibnamefont
  {Acu{\~n}a-Lara}}, \bibinfo {author} {\bibfnamefont {F.}~\bibnamefont
  {Sastre}},\ and\ \bibinfo {author} {\bibfnamefont {J.~R.}\ \bibnamefont
  {Vargas-Arriola}},\ }\bibfield  {title} {\bibinfo {title} {Critical phenomena
  in the majority voter model on two-dimensional regular lattices},\
  }\href@noop {} {\bibfield  {journal} {\bibinfo  {journal} {Physical Review
  E}\ }\textbf {\bibinfo {volume} {89}},\ \bibinfo {pages} {052109} (\bibinfo
  {year} {2014})}\BibitemShut {NoStop}%
\bibitem [{\citenamefont {Porter}\ \emph {et~al.}(2009)\citenamefont {Porter},
  \citenamefont {Onnela},\ and\ \citenamefont {Mucha}}]{porter2009communities}%
  \BibitemOpen
  \bibfield  {author} {\bibinfo {author} {\bibfnamefont {M.~A.}\ \bibnamefont
  {Porter}}, \bibinfo {author} {\bibfnamefont {J.-P.}\ \bibnamefont {Onnela}},\
  and\ \bibinfo {author} {\bibfnamefont {P.~J.}\ \bibnamefont {Mucha}},\
  }\bibfield  {title} {\bibinfo {title} {Communities in networks},\ }\href@noop
  {} {\bibfield  {journal} {\bibinfo  {journal} {Notices of the AMS}\ }\textbf
  {\bibinfo {volume} {56}},\ \bibinfo {pages} {1082} (\bibinfo {year}
  {2009})}\BibitemShut {NoStop}%
\bibitem [{\citenamefont {Fortunato}\ and\ \citenamefont
  {Hric}(2016)}]{fortunato2016community}%
  \BibitemOpen
  \bibfield  {author} {\bibinfo {author} {\bibfnamefont {S.}~\bibnamefont
  {Fortunato}}\ and\ \bibinfo {author} {\bibfnamefont {D.}~\bibnamefont
  {Hric}},\ }\bibfield  {title} {\bibinfo {title} {Community detection in
  networks: {A} user guide},\ }\href@noop {} {\bibfield  {journal} {\bibinfo
  {journal} {Physics Reports}\ }\textbf {\bibinfo {volume} {659}},\ \bibinfo
  {pages} {1} (\bibinfo {year} {2016})}\BibitemShut {NoStop}%
\bibitem [{\citenamefont {Holland}\ \emph {et~al.}(1983)\citenamefont
  {Holland}, \citenamefont {Laskey},\ and\ \citenamefont
  {Leinhardt}}]{holland1983stochastic}%
  \BibitemOpen
  \bibfield  {author} {\bibinfo {author} {\bibfnamefont {P.~W.}\ \bibnamefont
  {Holland}}, \bibinfo {author} {\bibfnamefont {K.~B.}\ \bibnamefont
  {Laskey}},\ and\ \bibinfo {author} {\bibfnamefont {S.}~\bibnamefont
  {Leinhardt}},\ }\bibfield  {title} {\bibinfo {title} {Stochastic blockmodels:
  First steps},\ }\href@noop {} {\bibfield  {journal} {\bibinfo  {journal}
  {Social Networks}\ }\textbf {\bibinfo {volume} {5}},\ \bibinfo {pages} {109}
  (\bibinfo {year} {1983})}\BibitemShut {NoStop}%
\bibitem [{\citenamefont {Peixoto}(2019)}]{peixoto2019bayesian}%
  \BibitemOpen
  \bibfield  {author} {\bibinfo {author} {\bibfnamefont {T.~P.}\ \bibnamefont
  {Peixoto}},\ }\bibfield  {title} {\bibinfo {title} {Bayesian stochastic
  blockmodeling},\ }in\ \href@noop {} {\emph {\bibinfo {booktitle} {Advances in
  Network Clustering and Blockmodeling}}},\ \bibinfo {editor} {edited by\
  \bibinfo {editor} {\bibfnamefont {A.~F.}\ \bibnamefont {Patrick~Doreian},
  \bibfnamefont {Vladimir~Batagelj}}}\ (\bibinfo  {publisher} {John Wiley \&
  Sons, Ltd, Hoboken, NJ},\ \bibinfo {year} {2019})\ pp.\ \bibinfo {pages}
  {289--332}\BibitemShut {NoStop}%
\bibitem [{\citenamefont {Lambiotte}\ and\ \citenamefont
  {Ausloos}(2007)}]{lambiotte2007coexistence}%
  \BibitemOpen
  \bibfield  {author} {\bibinfo {author} {\bibfnamefont {R.}~\bibnamefont
  {Lambiotte}}\ and\ \bibinfo {author} {\bibfnamefont {M.}~\bibnamefont
  {Ausloos}},\ }\bibfield  {title} {\bibinfo {title} {Coexistence of opposite
  opinions in a network with communities},\ }\href@noop {} {\bibfield
  {journal} {\bibinfo  {journal} {Journal of Statistical Mechanics: Theory and
  Experiment}\ }\textbf {\bibinfo {volume} {2007}},\ \bibinfo {pages} {P08026}
  (\bibinfo {year} {2007})}\BibitemShut {NoStop}%
\bibitem [{\citenamefont {Lambiotte}\ \emph {et~al.}(2007)\citenamefont
  {Lambiotte}, \citenamefont {Ausloos},\ and\ \citenamefont
  {Ho{\l}yst}}]{lambiotte2007majority}%
  \BibitemOpen
  \bibfield  {author} {\bibinfo {author} {\bibfnamefont {R.}~\bibnamefont
  {Lambiotte}}, \bibinfo {author} {\bibfnamefont {M.}~\bibnamefont {Ausloos}},\
  and\ \bibinfo {author} {\bibfnamefont {J.~A.}\ \bibnamefont {Ho{\l}yst}},\
  }\bibfield  {title} {\bibinfo {title} {Majority model on a network with
  communities},\ }\href@noop {} {\bibfield  {journal} {\bibinfo  {journal}
  {Physical Review E}\ }\textbf {\bibinfo {volume} {75}},\ \bibinfo {pages}
  {030101} (\bibinfo {year} {2007})}\BibitemShut {NoStop}%
\bibitem [{\citenamefont {Dasgupta}\ \emph {et~al.}(2009)\citenamefont
  {Dasgupta}, \citenamefont {Pan},\ and\ \citenamefont
  {Sinha}}]{dasgupta2009phase}%
  \BibitemOpen
  \bibfield  {author} {\bibinfo {author} {\bibfnamefont {S.}~\bibnamefont
  {Dasgupta}}, \bibinfo {author} {\bibfnamefont {R.~K.}\ \bibnamefont {Pan}},\
  and\ \bibinfo {author} {\bibfnamefont {S.}~\bibnamefont {Sinha}},\ }\bibfield
   {title} {\bibinfo {title} {Phase of {Ising} spins on modular networks
  analogous to social polarization},\ }\href@noop {} {\bibfield  {journal}
  {\bibinfo  {journal} {Physical Review E}\ }\textbf {\bibinfo {volume} {80}},\
  \bibinfo {pages} {025101} (\bibinfo {year} {2009})}\BibitemShut {NoStop}%
\bibitem [{\citenamefont {Bolfe}\ \emph {et~al.}(2018)\citenamefont {Bolfe},
  \citenamefont {Nicolao},\ and\ \citenamefont {Metz}}]{bolfe2018phase}%
  \BibitemOpen
  \bibfield  {author} {\bibinfo {author} {\bibfnamefont {M.}~\bibnamefont
  {Bolfe}}, \bibinfo {author} {\bibfnamefont {L.}~\bibnamefont {Nicolao}},\
  and\ \bibinfo {author} {\bibfnamefont {F.~L.}\ \bibnamefont {Metz}},\
  }\bibfield  {title} {\bibinfo {title} {Phase diagram and metastability of the
  {Ising} model on two coupled networks},\ }\href@noop {} {\bibfield  {journal}
  {\bibinfo  {journal} {Journal of Statistical Mechanics: Theory and
  Experiment}\ }\textbf {\bibinfo {volume} {2018}},\ \bibinfo {pages} {083404}
  (\bibinfo {year} {2018})}\BibitemShut {NoStop}%
\bibitem [{\citenamefont {Ru}\ and\ \citenamefont
  {Li-Ping}(2008)}]{ru2008opinion}%
  \BibitemOpen
  \bibfield  {author} {\bibinfo {author} {\bibfnamefont {W.}~\bibnamefont
  {Ru}}\ and\ \bibinfo {author} {\bibfnamefont {C.}~\bibnamefont {Li-Ping}},\
  }\bibfield  {title} {\bibinfo {title} {Opinion dynamics on complex networks
  with communities},\ }\href@noop {} {\bibfield  {journal} {\bibinfo  {journal}
  {Chinese Physics Letters}\ }\textbf {\bibinfo {volume} {25}},\ \bibinfo
  {pages} {1502} (\bibinfo {year} {2008})}\BibitemShut {NoStop}%
\bibitem [{\citenamefont {Liu}\ and\ \citenamefont
  {Hu}(2005)}]{liu2005epidemic}%
  \BibitemOpen
  \bibfield  {author} {\bibinfo {author} {\bibfnamefont {Z.}~\bibnamefont
  {Liu}}\ and\ \bibinfo {author} {\bibfnamefont {B.}~\bibnamefont {Hu}},\
  }\bibfield  {title} {\bibinfo {title} {Epidemic spreading in community
  networks},\ }\href@noop {} {\bibfield  {journal} {\bibinfo  {journal}
  {Europhysics Letters}\ }\textbf {\bibinfo {volume} {72}},\ \bibinfo {pages}
  {315} (\bibinfo {year} {2005})}\BibitemShut {NoStop}%
\bibitem [{\citenamefont {Wu}\ and\ \citenamefont
  {Liu}(2008)}]{wu2008community}%
  \BibitemOpen
  \bibfield  {author} {\bibinfo {author} {\bibfnamefont {X.}~\bibnamefont
  {Wu}}\ and\ \bibinfo {author} {\bibfnamefont {Z.}~\bibnamefont {Liu}},\
  }\bibfield  {title} {\bibinfo {title} {How community structure influences
  epidemic spread in social networks},\ }\href@noop {} {\bibfield  {journal}
  {\bibinfo  {journal} {Physica A: Statistical Mechanics and its Applications}\
  }\textbf {\bibinfo {volume} {387}},\ \bibinfo {pages} {623} (\bibinfo {year}
  {2008})}\BibitemShut {NoStop}%
\bibitem [{\citenamefont {Stegehuis}\ \emph {et~al.}(2016)\citenamefont
  {Stegehuis}, \citenamefont {Van Der~Hofstad},\ and\ \citenamefont
  {Van~Leeuwaarden}}]{stegehuis2016epidemic}%
  \BibitemOpen
  \bibfield  {author} {\bibinfo {author} {\bibfnamefont {C.}~\bibnamefont
  {Stegehuis}}, \bibinfo {author} {\bibfnamefont {R.}~\bibnamefont {Van
  Der~Hofstad}},\ and\ \bibinfo {author} {\bibfnamefont {J.~S.}\ \bibnamefont
  {Van~Leeuwaarden}},\ }\bibfield  {title} {\bibinfo {title} {Epidemic
  spreading on complex networks with community structures},\ }\href@noop {}
  {\bibfield  {journal} {\bibinfo  {journal} {Scientific Reports}\ }\textbf
  {\bibinfo {volume} {6}},\ \bibinfo {pages} {29748} (\bibinfo {year}
  {2016})}\BibitemShut {NoStop}%
\bibitem [{\citenamefont {Huang}\ \emph {et~al.}(2006)\citenamefont {Huang},
  \citenamefont {Park},\ and\ \citenamefont {Lai}}]{huang2006information}%
  \BibitemOpen
  \bibfield  {author} {\bibinfo {author} {\bibfnamefont {L.}~\bibnamefont
  {Huang}}, \bibinfo {author} {\bibfnamefont {K.}~\bibnamefont {Park}},\ and\
  \bibinfo {author} {\bibfnamefont {Y.-C.}\ \bibnamefont {Lai}},\ }\bibfield
  {title} {\bibinfo {title} {Information propagation on modular networks},\
  }\href@noop {} {\bibfield  {journal} {\bibinfo  {journal} {Physical Review
  E}\ }\textbf {\bibinfo {volume} {73}},\ \bibinfo {pages} {035103} (\bibinfo
  {year} {2006})}\BibitemShut {NoStop}%
\bibitem [{\citenamefont {Nematzadeh}\ \emph {et~al.}(2014)\citenamefont
  {Nematzadeh}, \citenamefont {Ferrara}, \citenamefont {Flammini},\ and\
  \citenamefont {Ahn}}]{nematzadeh2014optimal}%
  \BibitemOpen
  \bibfield  {author} {\bibinfo {author} {\bibfnamefont {A.}~\bibnamefont
  {Nematzadeh}}, \bibinfo {author} {\bibfnamefont {E.}~\bibnamefont {Ferrara}},
  \bibinfo {author} {\bibfnamefont {A.}~\bibnamefont {Flammini}},\ and\
  \bibinfo {author} {\bibfnamefont {Y.-Y.}\ \bibnamefont {Ahn}},\ }\bibfield
  {title} {\bibinfo {title} {Optimal network modularity for information
  diffusion},\ }\href@noop {} {\bibfield  {journal} {\bibinfo  {journal}
  {Physical Review Letters}\ }\textbf {\bibinfo {volume} {113}},\ \bibinfo
  {pages} {088701} (\bibinfo {year} {2014})}\BibitemShut {NoStop}%
\bibitem [{\citenamefont {Suchecki}\ and\ \citenamefont
  {Ho{\l}yst}(2009)}]{suchecki2009bistable}%
  \BibitemOpen
  \bibfield  {author} {\bibinfo {author} {\bibfnamefont {K.}~\bibnamefont
  {Suchecki}}\ and\ \bibinfo {author} {\bibfnamefont {J.~A.}\ \bibnamefont
  {Ho{\l}yst}},\ }\bibfield  {title} {\bibinfo {title} {Bistable-monostable
  transition in the {Ising} model on two connected complex networks},\
  }\href@noop {} {\bibfield  {journal} {\bibinfo  {journal} {Physical Review
  E}\ }\textbf {\bibinfo {volume} {80}},\ \bibinfo {pages} {031110} (\bibinfo
  {year} {2009})}\BibitemShut {NoStop}%
\bibitem [{\citenamefont {Kivel{\"a}}\ \emph {et~al.}(2014)\citenamefont
  {Kivel{\"a}}, \citenamefont {Arenas}, \citenamefont {Barthelemy},
  \citenamefont {Gleeson}, \citenamefont {Moreno},\ and\ \citenamefont
  {Porter}}]{kivela2014multilayer}%
  \BibitemOpen
  \bibfield  {author} {\bibinfo {author} {\bibfnamefont {M.}~\bibnamefont
  {Kivel{\"a}}}, \bibinfo {author} {\bibfnamefont {A.}~\bibnamefont {Arenas}},
  \bibinfo {author} {\bibfnamefont {M.}~\bibnamefont {Barthelemy}}, \bibinfo
  {author} {\bibfnamefont {J.~P.}\ \bibnamefont {Gleeson}}, \bibinfo {author}
  {\bibfnamefont {Y.}~\bibnamefont {Moreno}},\ and\ \bibinfo {author}
  {\bibfnamefont {M.~A.}\ \bibnamefont {Porter}},\ }\bibfield  {title}
  {\bibinfo {title} {Multilayer networks},\ }\href@noop {} {\bibfield
  {journal} {\bibinfo  {journal} {Journal of Complex Networks}\ }\textbf
  {\bibinfo {volume} {2}},\ \bibinfo {pages} {203} (\bibinfo {year}
  {2014})}\BibitemShut {NoStop}%
\bibitem [{\citenamefont {Aleta}\ and\ \citenamefont
  {Moreno}(2019)}]{aleta2019multilayer}%
  \BibitemOpen
  \bibfield  {author} {\bibinfo {author} {\bibfnamefont {A.}~\bibnamefont
  {Aleta}}\ and\ \bibinfo {author} {\bibfnamefont {Y.}~\bibnamefont {Moreno}},\
  }\bibfield  {title} {\bibinfo {title} {Multilayer networks in a nutshell},\
  }\href@noop {} {\bibfield  {journal} {\bibinfo  {journal} {Annual Review of
  Condensed Matter Physics}\ }\textbf {\bibinfo {volume} {10}},\ \bibinfo
  {pages} {45} (\bibinfo {year} {2019})}\BibitemShut {NoStop}%
\bibitem [{\citenamefont {{De Domenico}}(2022)}]{de2022multilayer}%
  \BibitemOpen
  \bibfield  {author} {\bibinfo {author} {\bibfnamefont {M.}~\bibnamefont {{De
  Domenico}}},\ }\href@noop {} {\emph {\bibinfo {title} {Multilayer Networks:
  Analysis and Visualization}}}\ (\bibinfo  {publisher} {Springer, Cham,
  Switzerland},\ \bibinfo {year} {2022})\BibitemShut {NoStop}%
\bibitem [{\citenamefont {Diakonova}\ \emph {et~al.}(2016)\citenamefont
  {Diakonova}, \citenamefont {Nicosia}, \citenamefont {Latora},\ and\
  \citenamefont {San~Miguel}}]{diakonova2016irreducibility}%
  \BibitemOpen
  \bibfield  {author} {\bibinfo {author} {\bibfnamefont {M.}~\bibnamefont
  {Diakonova}}, \bibinfo {author} {\bibfnamefont {V.}~\bibnamefont {Nicosia}},
  \bibinfo {author} {\bibfnamefont {V.}~\bibnamefont {Latora}},\ and\ \bibinfo
  {author} {\bibfnamefont {M.}~\bibnamefont {San~Miguel}},\ }\bibfield  {title}
  {\bibinfo {title} {Irreducibility of multilayer network dynamics: The case of
  the voter model},\ }\href@noop {} {\bibfield  {journal} {\bibinfo  {journal}
  {New Journal of Physics}\ }\textbf {\bibinfo {volume} {18}},\ \bibinfo
  {pages} {023010} (\bibinfo {year} {2016})}\BibitemShut {NoStop}%
\bibitem [{\citenamefont {Krawiecki}\ \emph {et~al.}(2018)\citenamefont
  {Krawiecki}, \citenamefont {Gradowski},\ and\ \citenamefont
  {Siudem}}]{krawiecki2018majority}%
  \BibitemOpen
  \bibfield  {author} {\bibinfo {author} {\bibfnamefont {A.}~\bibnamefont
  {Krawiecki}}, \bibinfo {author} {\bibfnamefont {T.}~\bibnamefont
  {Gradowski}},\ and\ \bibinfo {author} {\bibfnamefont {G.}~\bibnamefont
  {Siudem}},\ }\bibfield  {title} {\bibinfo {title} {Majority vote model on
  multiplex networks},\ }\href@noop {} {\bibfield  {journal} {\bibinfo
  {journal} {Acta Physica Polonica A}\ }\textbf {\bibinfo {volume} {133}},\
  \bibinfo {pages} {1433} (\bibinfo {year} {2018})}\BibitemShut {NoStop}%
\bibitem [{\citenamefont {Choi}\ and\ \citenamefont
  {Goh}(2019)}]{choi2019majority}%
  \BibitemOpen
  \bibfield  {author} {\bibinfo {author} {\bibfnamefont {J.}~\bibnamefont
  {Choi}}\ and\ \bibinfo {author} {\bibfnamefont {K.-I.}\ \bibnamefont {Goh}},\
  }\bibfield  {title} {\bibinfo {title} {Majority-vote dynamics on multiplex
  networks with two layers},\ }\href@noop {} {\bibfield  {journal} {\bibinfo
  {journal} {New Journal of Physics}\ }\textbf {\bibinfo {volume} {21}},\
  \bibinfo {pages} {035005} (\bibinfo {year} {2019})}\BibitemShut {NoStop}%
\bibitem [{\citenamefont {Chmiel}\ \emph {et~al.}(2017)\citenamefont {Chmiel},
  \citenamefont {Sienkiewicz},\ and\ \citenamefont
  {Sznajd-Weron}}]{chmiel2017tricriticality}%
  \BibitemOpen
  \bibfield  {author} {\bibinfo {author} {\bibfnamefont {A.}~\bibnamefont
  {Chmiel}}, \bibinfo {author} {\bibfnamefont {J.}~\bibnamefont
  {Sienkiewicz}},\ and\ \bibinfo {author} {\bibfnamefont {K.}~\bibnamefont
  {Sznajd-Weron}},\ }\bibfield  {title} {\bibinfo {title} {Tricriticality in
  the $q$-neighbor {Ising} model on a partially duplex clique},\ }\href@noop {}
  {\bibfield  {journal} {\bibinfo  {journal} {Physical Review E}\ }\textbf
  {\bibinfo {volume} {96}},\ \bibinfo {pages} {062137} (\bibinfo {year}
  {2017})}\BibitemShut {NoStop}%
\bibitem [{\citenamefont {Gradowski}\ and\ \citenamefont
  {Krawiecki}(2020)}]{gradowski2020pair}%
  \BibitemOpen
  \bibfield  {author} {\bibinfo {author} {\bibfnamefont {T.}~\bibnamefont
  {Gradowski}}\ and\ \bibinfo {author} {\bibfnamefont {A.}~\bibnamefont
  {Krawiecki}},\ }\bibfield  {title} {\bibinfo {title} {Pair approximation for
  the $q$-voter model with independence on multiplex networks},\ }\href@noop {}
  {\bibfield  {journal} {\bibinfo  {journal} {Physical Review E}\ }\textbf
  {\bibinfo {volume} {102}},\ \bibinfo {pages} {022314} (\bibinfo {year}
  {2020})}\BibitemShut {NoStop}%
\bibitem [{\citenamefont {Lee}\ \emph {et~al.}(2014)\citenamefont {Lee},
  \citenamefont {Brummitt},\ and\ \citenamefont {Goh}}]{lee2014threshold}%
  \BibitemOpen
  \bibfield  {author} {\bibinfo {author} {\bibfnamefont {K.-M.}\ \bibnamefont
  {Lee}}, \bibinfo {author} {\bibfnamefont {C.~D.}\ \bibnamefont {Brummitt}},\
  and\ \bibinfo {author} {\bibfnamefont {K.-I.}\ \bibnamefont {Goh}},\
  }\bibfield  {title} {\bibinfo {title} {Threshold cascades with response
  heterogeneity in multiplex networks},\ }\href@noop {} {\bibfield  {journal}
  {\bibinfo  {journal} {Physical Review E}\ }\textbf {\bibinfo {volume} {90}},\
  \bibinfo {pages} {062816} (\bibinfo {year} {2014})}\BibitemShut {NoStop}%
\bibitem [{\citenamefont {Nguyen}\ \emph {et~al.}(2018)\citenamefont {Nguyen},
  \citenamefont {Xiao}, \citenamefont {Xu}, \citenamefont {Li},\ and\
  \citenamefont {Wang}}]{nguyen2018opinion}%
  \BibitemOpen
  \bibfield  {author} {\bibinfo {author} {\bibfnamefont {V.~X.}\ \bibnamefont
  {Nguyen}}, \bibinfo {author} {\bibfnamefont {G.}~\bibnamefont {Xiao}},
  \bibinfo {author} {\bibfnamefont {X.-J.}\ \bibnamefont {Xu}}, \bibinfo
  {author} {\bibfnamefont {G.}~\bibnamefont {Li}},\ and\ \bibinfo {author}
  {\bibfnamefont {Z.}~\bibnamefont {Wang}},\ }\bibfield  {title} {\bibinfo
  {title} {Opinion formation on multiplex scale-free networks},\ }\href@noop {}
  {\bibfield  {journal} {\bibinfo  {journal} {Europhysics Letters}\ }\textbf
  {\bibinfo {volume} {121}},\ \bibinfo {pages} {26002} (\bibinfo {year}
  {2018})}\BibitemShut {NoStop}%
\bibitem [{\citenamefont {Bazzi}\ \emph {et~al.}(2020)\citenamefont {Bazzi},
  \citenamefont {Jeub}, \citenamefont {Arenas}, \citenamefont {Howison},\ and\
  \citenamefont {Porter}}]{bazzi2020framework}%
  \BibitemOpen
  \bibfield  {author} {\bibinfo {author} {\bibfnamefont {M.}~\bibnamefont
  {Bazzi}}, \bibinfo {author} {\bibfnamefont {L.~G.}\ \bibnamefont {Jeub}},
  \bibinfo {author} {\bibfnamefont {A.}~\bibnamefont {Arenas}}, \bibinfo
  {author} {\bibfnamefont {S.~D.}\ \bibnamefont {Howison}},\ and\ \bibinfo
  {author} {\bibfnamefont {M.~A.}\ \bibnamefont {Porter}},\ }\bibfield  {title}
  {\bibinfo {title} {A framework for the construction of generative models for
  mesoscale structure in multilayer networks},\ }\href@noop {} {\bibfield
  {journal} {\bibinfo  {journal} {Physical Review Research}\ }\textbf {\bibinfo
  {volume} {2}},\ \bibinfo {pages} {023100} (\bibinfo {year}
  {2020})}\BibitemShut {NoStop}%
\bibitem [{\citenamefont {Gleeson}\ \emph {et~al.}(2012)\citenamefont
  {Gleeson}, \citenamefont {Melnik}, \citenamefont {Ward}, \citenamefont
  {Porter},\ and\ \citenamefont {Mucha}}]{gleeson2012accuracy}%
  \BibitemOpen
  \bibfield  {author} {\bibinfo {author} {\bibfnamefont {J.~P.}\ \bibnamefont
  {Gleeson}}, \bibinfo {author} {\bibfnamefont {S.}~\bibnamefont {Melnik}},
  \bibinfo {author} {\bibfnamefont {J.~A.}\ \bibnamefont {Ward}}, \bibinfo
  {author} {\bibfnamefont {M.~A.}\ \bibnamefont {Porter}},\ and\ \bibinfo
  {author} {\bibfnamefont {P.~J.}\ \bibnamefont {Mucha}},\ }\bibfield  {title}
  {\bibinfo {title} {Accuracy of mean-field theory for dynamics on real-world
  networks},\ }\href@noop {} {\bibfield  {journal} {\bibinfo  {journal}
  {Physical Review E}\ }\textbf {\bibinfo {volume} {85}},\ \bibinfo {pages}
  {026106} (\bibinfo {year} {2012})}\BibitemShut {NoStop}%
\bibitem [{\citenamefont {Peng}(2022)}]{code}%
  \BibitemOpen
  \bibfield  {author} {\bibinfo {author} {\bibfnamefont {K.}~\bibnamefont
  {Peng}},\ }\href@noop {} {\bibinfo {title} {Majority-vote model on multiplex
  networks with community structure}},\ \bibinfo {howpublished}
  {https://gitlab.com/KaiyanP196/majority-vote-model-on-multiplex-networks-with-community-structure}
  (\bibinfo {year} {2022})\BibitemShut {NoStop}%
\bibitem [{Note1()}]{Note1}%
  \BibitemOpen
  \bibinfo {note} {\protect \textsc {Matlab}'s {\protect \texttt contour}
  function finds isolines of a matrix through interpolation.}\BibitemShut
  {Stop}%
\end{thebibliography}


%


\end{document}